\documentclass[twocolumn]{aastex62}
\usepackage{amsmath}
\newcommand{\package}[1]{\textsl{#1}}

\renewcommand{\tablename}{Table}

\shorttitle{Streams in Cen A}
\shortauthors{Pearson et al.}

\begin{document}\sloppy\sloppypar\raggedbottom\frenchspacing 

\title{Mapping Dark Matter with Extragalactic Stellar Streams:\\ the Case of Centaurus A}

 \author[0000-0003-0256-5446]{Sarah Pearson}\thanks{Hubble Fellow}
\affiliation{Center for Cosmology and Particle Physics, Department of Physics, New York University, 726 Broadway, New York, NY 10003, USA}
 \email{spearson@nyu.edu}
 \correspondingauthor{Sarah Pearson}

\author[0000-0003-0872-7098]{Adrian~M.~Price-Whelan}
\affiliation{Center for Computational Astrophysics, Flatiron Institute, 162 5th Ave, New York City, NY 10010, USA}

\author[0000-0003-2866-9403]{David W. Hogg}
\affiliation{Center for Cosmology and Particle Physics, Department of Physics, New York University, 726 Broadway, New York, NY 10003, USA}
\affiliation{Center for Computational Astrophysics, Flatiron Institute, 162 5th Ave, New York City, NY 10010, USA}
\affiliation{Max-Planck-Institut f\"ur Astronomie, K\"onigstuhl 17, D-69117 Heidelberg, Germany}

\author[0000-0003-0248-5470]{Anil C. Seth}
 \affiliation{Department of Physics and Astronomy, University of Utah, Salt Lake City, UT 84112, USA}

\author[0000-0003-4102-380X]{David J. Sand}
 \affiliation{Department of Astronomy and Steward Observatory, University of Arizona, 933 N Cherry Ave, Tucson, AZ 85719, USA}

 \author[0000-0001-8917-1532]{Jason A. S. Hunt}
\affiliation{Center for Computational Astrophysics, Flatiron Institute, 162 5th Ave, New York City, NY 10010, USA}

\author[0000-0002-1763-4128]{Denija Crnojevi\'{c}}
\affil{University of Tampa, 401 West Kennedy Boulevard, Tampa, FL 33606, USA}

\begin{abstract}\noindent 
In the coming decade, thousands of stellar streams will be observed in the halos of external galaxies.
What fundamental discoveries will we make about dark matter from these streams?
As a first attempt to look at these questions, we model Magellan/Megacam imaging of the Centaurus A's (Cen~A) disrupting dwarf companion Dwarf 3 (Dw3) and its associated stellar stream, to find out what can be learned about the Cen~A dark-matter halo.
We develop a novel external galaxy stream-fitting technique and generate model stellar streams that reproduce the stream morphology visible in the imaging.
We find that there are many viable stream models that fit the data well, with reasonable parameters,
provided that Cen~A has a halo mass larger than M$_{200}$ $>4.70\times 10^{12}$ M$_{\odot}$. There is a second stream in Cen A's halo that is also reproduced within the context of this same dynamical model. 
However, stream morphology in the imaging alone does not uniquely determine the mass or mass distribution for the Cen~A halo.
In particular, the stream models with high likelihood show covariances between the inferred Cen~A mass distribution, the inferred Dw3 progenitor mass, the Dw3 velocity, and the Dw3 line-of-sight position.
We show that these degeneracies can be broken with radial-velocity measurements along the stream, and that a single radial velocity measurement puts a substantial lower limit on the halo mass.
These results suggest that targeted radial-velocity measurements will be critical if we want to learn about dark matter from extragalactic stellar streams.
\end{abstract}


\section{Introduction} \label{sec:intro}
When a dwarf galaxy is accreted by a larger galaxy, a tidal interaction unfolds that can lead to the formation of stellar streams \citep[e.g.,][]{Johnston1995}. The collective velocities and positions of stars in stellar streams can store dynamical information from billions of years of past evolution \citep[e.g.,][]{Johnston2001}. 

From studies in the Milky Way (MW), we know that stellar streams provide information on the Galactic accretion history \citep[e.g.,][]{belokurov2006,Naidu2020}. Streams in the MW have also been used to constrain the enclosed dark matter distribution of our Galaxy \citep[e.g.,][]{koposov2010,kuepper2015}, and to investigate the shape of the MW's dark matter halo \citep[e.g.,][]{law2010,veraciro2013,pearson2015,Bovy2016}. Additionally, stellar streams trace out orbits of co-moving stars close in energy and angular momentum space, which make streams powerful probes of orbit structures in galaxies \citep[i.e, thin streams can only exist on regular or near-resonant orbits:][Yavetz et al., in prep.]{price2016,Yavetz2021}.

\citet{Bonaca2018} showed that stream tracks in the MW hold key information on the local acceleration  field and enclosed  mass, and that longer streams on more eccentric orbits best constrain the dark matter halo shape. They also show that the joint constraint, from multiple streams, provides more information than the sum of individual stream tracks. 
In the MW, we can obtain 6D phase-space measurements 
for individual stars in stellar streams \citep[e.g.,][]{fritz2015,price2018,Shipp2019,li2019,Li2021}.
However, in external galaxies we can not directly build on the intuition from \citet{Bonaca2018}, since we observe streams in projection, where the physical scale of the system depends on the distance to the external galaxy of interest. We often only have access to 2D morphological measurements, but stellar kinematic and distance measurements do exist for some streams (e.g., from GCs:  \citealt{Veljanoski2014}, from surface brightness fluctuations: \citealt{toloba2016}, and from HST observations of the tip of the red-giant branch: \citealt{denja:2019}).

Individual stellar streams from accreted dwarf galaxies have been observed around both massive external galaxies \citep[e.g.,][]{shang1998,martin2014,delgado2021} and around dwarf galaxies \citep[e.g.,][]{delgado2012,toloba2016,fong2018,carlin2019,fong2020}. We are finally entering an era in astronomy, where we will have access to statistical samples of stellar streams that orbit external galaxies. Ground-based surveys, such as the Vera Rubin Observatory (Rubin; \citealt{Ivezic08}), and space telescopes, such as Euclid (\citealt{laureijs11}), and the Nancy Grace Roman Space Telescope (Roman; \citealt{spergel13}), will discover thousands of stellar streams in galaxies out to hundreds of Mpc \citep[e.g.,][]{Laine2018}.
The astronomical community has prepared for this by developing techniques to automate stellar stream detection in external galaxies \citep[e.g.,][]{pearson2021, Pearson2022}, and to automatically classify substructure \citep[e.g., streams vs. shells in][]{Hendel2019}. 

While stellar streams in external galaxies could provide a fundamental way of mapping dark matter, we still lack theoretical counterparts to interpret stellar stream observations.
\citet{Fardal2013} modeled the Giant
Southern Stream (GSS) and used Bayesian sampling of Nbody simulations to infer M31's mass \citep[see also][]{fardal2007,fardal2009}. They took advantage of data from both photometric mapping projects such as PAndAS \citep[e.g.,][]{McConnachie2009,McConnachie2018} and of radial velocities measurements for individual red giant branch stars in the halo of M31 which were obtained with 8-10m class telescopes \citep[e.g.,][]{ibata2004,gilbert2009}. The best fit model in \citet{Fardal2013} reproduce both the observed distance and velocity gradients along the stream. But in most external galaxies, we do not have access to such info. \citet{dokkum2019} provided an example of a fit to a stream well beyond the local group, NGC 5907. While they  showed that a match to the stream morphology was possible, they did not explore the stream's constraining power on NGC 5907's halo parameters. 
It therefore remains largely unexplored what we can learn about external galaxies' dark matter halo properties  beyond the local group (i.e., mass distributions, dark matter concentrations, and shapes) from their populations of stellar streams.

In this paper, we take the first step towards using extragalactic stellar streams beyond the local group to map dark matter. Specifically, we explore what the streams around the elliptical galaxy, Centaurus A (Cen A), located 3.8 Mpc \citep{Harris2010} from the Milky Way, can teach us about Cen A's dark matter halo. We first develop a stream-modeling technique for streams evolved in external galaxies, and compare model streams to data from the Magellan Clay 6.5m telescope \citep{denja:2016}. This method can be applied to any external galaxy. 
We find that we can easily reproduce the morphology of the Dw3 stream in an NFW potential motivated by the enclosed mass of globular clusters (GCs) in Cen A \citep{Woodley2010}. We also find that the inferred halo mass of Cen A is degenerate with stream morphology if we do not include any radial velocity measurements of Dw3. When we include a radial velocity, however, the Dw3 stream in Cen A prefers a halo mass of at least M$_{200}$ $>$ $4.70 \times10^{12}$ M$_{\odot}$. We demonstrate that radial velocities and distance measurements along the stream can help constrain the dark matter halo mass further as expected from stream modeling efforts in the MW. 

The paper is organized as follows: In Section \ref{sec:data}, we present the data of Cen A. In Section \ref{sec:methods} we first present our method for generating model streams in Cen A's potential and our assessment of how well the model streams fit the data. In Section \ref{sec:results}, we present the results of our analysis. In Section \ref{sec:discussion} we discuss our results, and we conclude in Section \ref{sec:conclusion}.

\section{Data}\label{sec:data}
\begin{figure*}
\centerline{\includegraphics[width=\textwidth]{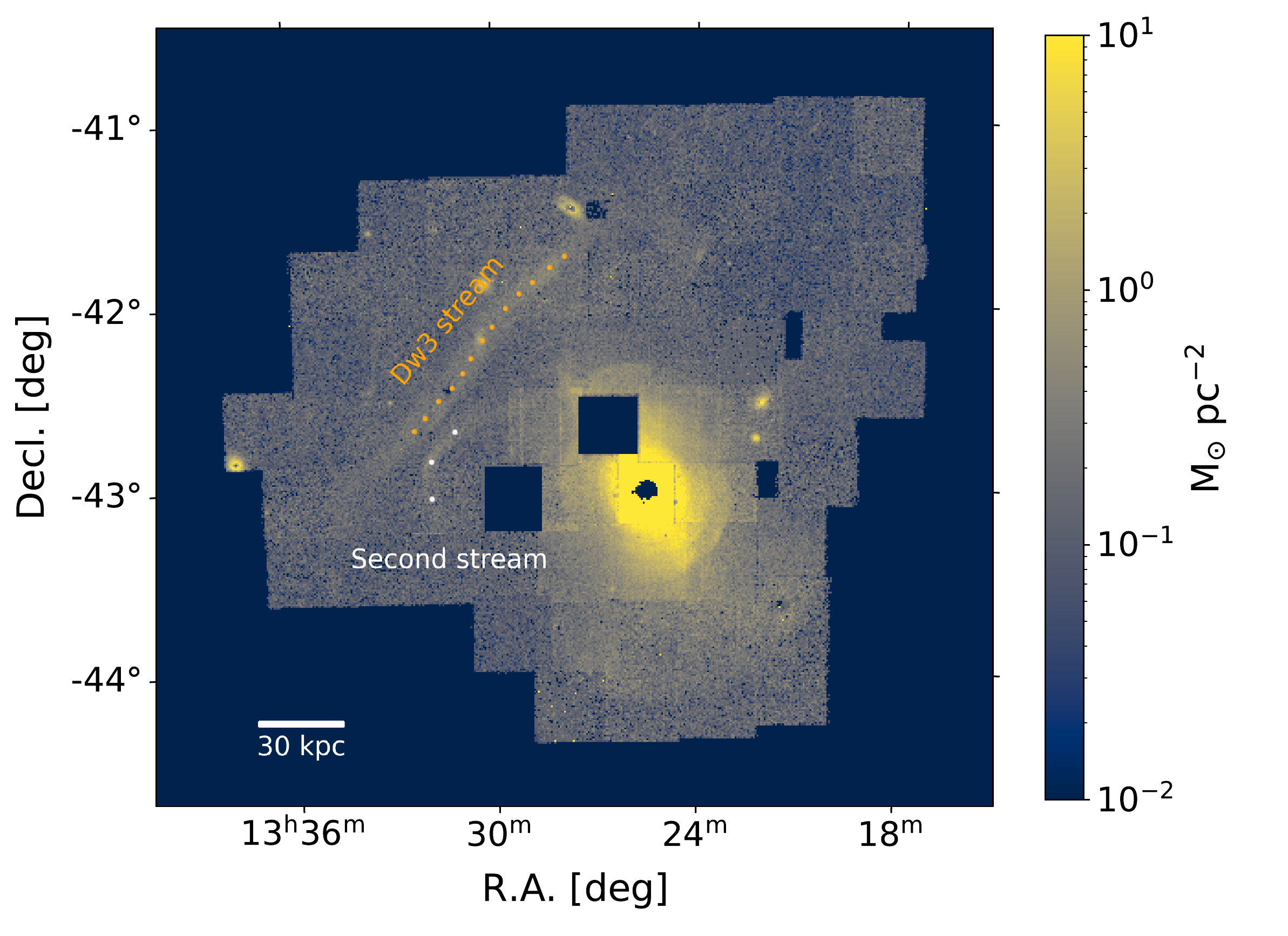}}
\caption{The outer stellar halo density map of RGB stars from \citet{denja:2016} obtained with the Magellan Clay 6.5m telescope calibrated to be in units of M$_{\odot}{\rm pc}^{-2}$. We have overplotted 13 control data stream center points (black) along the Dw3 Stream and 3 control points (white) along the second stream in Cen A. Note that the actual Cen A galaxy is very small compared to the stellar halo mapped here. 
}
\label{fig:cena_data}
\end{figure*}

\begin{deluxetable*}{lccccccc}
\tablecaption{Observational data used for Dwarf 3 stream modeling}
\tablecolumns{8}
\tablenum{1}\label{tab:cena}
\tablewidth{0pt}
\tablehead{\colhead{Parameters} & 
\colhead{R.A. } &
\colhead{decl.} &
\colhead{M$_{\rm dynamical}$} &
\colhead{L$_K$} &
\colhead{r$_h$} &
\colhead{distance} &
\colhead{vel$_{\rm rad}$}\\
\colhead{} & 
\colhead{[deg]} &
\colhead{[deg]} &
\colhead{[M$_{\odot}$]} &
\colhead{[L$_{\odot}$]} &
\colhead{[kpc]} &
\colhead{[Mpc]} &
\colhead{[km s$^{-1}$]} 
}

\startdata
Cen A      &201.365063&$-43.019113$ &  $9\times 10^{11}$\tablenotemark{,a}& $1.5\times10^{11}$$^{,b}$ & &3.8 $\pm~ 0.1$$^{c}$ & 541 $\pm~ 7$$^{d}$ \\
Dwarf 3$^{e}$ &202.585167&   $-42.191741$   & -& $1.5\times 10^7$& 2.49 $\pm~ 0.17$ &3.88 $\pm~ 0.16$& 359.6  $\pm~ 2.4$$^{f}$\\
CenA/Dw3$_{\rm relative}$&-&-    & - & - &- &0.079$^{g}$ &$-181.4$ \\
Second stream & 202.975 & $-42.85$ &-&-&-&3.8 $^h$ & 651 $\pm 13^i$ 
\enddata
\tablenotetext{a}{From GCs within 40 kpc \citep{Woodley2010}, $^b$: \citet{karachentsev2002}, $^c$: \citet{Harris2010}, $^d$: From planetary nebulae \citet{hui1995}, $^e$: See also \citet{denja:2019}, $^f$: \citet{dumont2022}, $^g$: Projected distance (equivalent to 1.22 degrees) assuming a Cen A distance of 3.8 Mpc, $^h$: We place the second stream at 3.8 Mpc, since \citet{denja:2016} find that this stream and Cen A are at the same distance. $^i$: \citet{Hughes2022}.}
\end{deluxetable*} 
Centaurus A (Cen A or NGC 5128) is located 3.8 Mpc from the Milky Way \citep{Harris2010}, and is the nearest massive early type galaxy. The galaxy has been observed extensively at a range of wavelengths \citep[e.g.,][]{Graham1979,fabbiano1992,Harris99}, and it has a pronounced dust ring near its center \citep{jarret2003}. Its stellar luminosity has been measured to be $L_K = 1.5\times10^{11}$ L$_{\odot}$ \citep{karachentsev2002}. With a mass-to-light ratio of 0.7 \citep{silge05}, this corresponds to a stellar mass of $M_* = 1 \times 10^{11}$ M$_{\odot}$.  Several groups have analysed its diffuse light \citep{denja:2016}, globular cluster populations \citep[e.g.][]{Woodley2010,hughes2021,dumont2022}, satellite populations \citep{denja:2019,mueller2022} as well as its stellar streams \citep{denja:2016,denja:2019}. 

In this paper, we use Magellan/Megacam imaging data  \citep[][]{McLeod2015} from the Magellan Clay 6.5m telescope, Las Campanas Observatory \citep[][]{denja:2016}. We present an RGB stellar density map calibrated to be in units of M$_{\odot}{\rm pc}^{-2}$ in Figure \ref{fig:cena_data}.  This calibration was done using 6 HST pointings associated with Dw3 and its stream 
from \citet{denja:2019} combined with Padova isochrones \citep{Bressan2012} to convert number counts of RGB stars in the fields into stellar mass for a range of metallicities and ages.  We averaged the results across the six fields -- the derived conversion factor varies by a factor of $\sim$2 between fields, this scatter exceeds the model variations with metallicity and age.  
We mark the location of the Dwarf 3 stream as well as the ``second stream.'' Throughout this work, we also compare our simulated streams to the measured relative distance along the stream, obtained from modeling the stellar populations in Hubble Space Telescope imaging of the Dw3 stream \citep{denja:2019}. We additionally use the radial velocity measurement of the nuclear star cluster in Dw3 from \citet{dumont2022}, and the radial velocity measurement of Cen A from \citet{hui1995} and \citet{hughes2021}. In Table \ref{tab:cena}, we summarize the observational constraints that we use throughout the paper for Cen A and its streams.

At present day, the Dw3 luminosity is L$_V = 1.5 \times 10^{7}$ L$_{\odot}$ excluding its stellar stream. If we assume a mass-to-light ratio of 2, which is appropriate for an old, relatively metal poor stellar population, this means that the present day stellar mass is $m_{\rm prog} \sim 3 \times 10^7$ M$_{\odot}$. 
In \citet{denja:2019}, they estimate the original luminosity, using MegaCam data and assuming a distance of 3.88 Mpc, to be M$_V \sim -15$. This corresponds to L$_V = 8\times 10^{7}$ L$_{\odot}$, and thus an initial Dw3 mass of $m_{\rm prog} \sim 1.6\times10^8$ M$_{\odot}$, again assuming a mass-to-light ratio of 2. Throughout the paper, we focus on the stage where the stars are being stripped and ignore any initial dark matter in the progenitor (see a further discussion of this in Section \ref{sec:nbodyfollowup} and \ref{sec:nbody}).

\section{Methods}\label{sec:methods}

Our goal is to fit the on-sky track of the Dw3 stream and radial velocity of the Dw3 progenitor to determine what we can learn about the dark matter distribution in Cen A. 
In this Section, we first present the potential we use to simulate Dw3's evolution in the Cen A halo (Section \ref{sec:potential}), we then present the control data points and coordinate system used in the modeling (Section \ref{sec:control}), followed by a description of our external stream generator (Section \ref{sec:streamgenerator}) and our stream-fitting technique (Section \ref{sec:streamfit}). We lastly explain how we run follow-up N-body simulations of our best-fit Dw3 stream in Cen A's halo (Section \ref{sec:nbodyfollowup}).

\subsection{Potentials}\label{sec:potential}
\begin{figure}
\centerline{\includegraphics[width=\columnwidth]{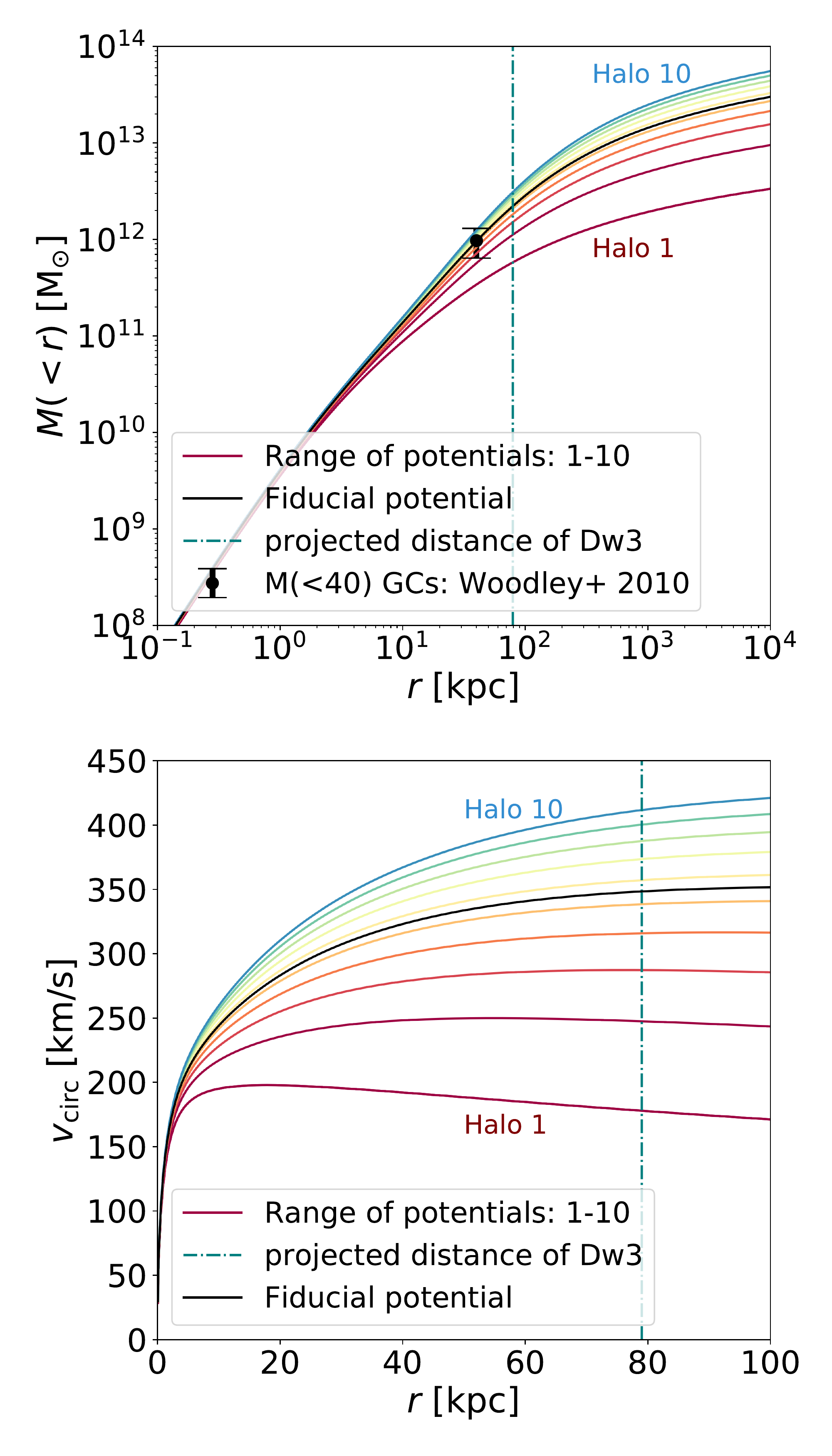}}
\caption{Top: Enclosed mass of the fiducial potential as a function of galactic radius (black solid line). This potential includes a Hernquist stellar component and an NFW dark matter halo component (M$_{\rm 200,fiducial} = 9.4 \times 10^{12}$ M$_{\odot}$) motivated from GC kinematic estimates within 40 kpc \citep[][see black data point]{Woodley2010}. The colored lines shows a range of potentials with the same Hernquist stellar component, but with varying NFW profiles (colored lines; M$_{200} = 0.94 - 17.86~ \times 10^{12}$ M$_{\odot}$, where red is Halo 1 and blue is Halo 10), which we use throughout the paper (see also Table \ref{tab:pot}). We use the halo mass vs halo concentration relation from \citet{Munoz2011}. The vertical teal line shows the projected distance from Dw3 to the center of Cen A. Bottom: Circular velocity as a function of galactic radius for the fiducial potential (black line) as well as the range of potentials (colored lines). 
}
\label{fig:potential}
\end{figure}

We use a two-component mass model to represent the gravitational field of Cen A,  with a Hernquist spherical stellar component with M$_* = 1.5\times 10^{11}$ M$_{\odot}$ and a concentration b = 4.896 kpc  \citep{hernquist1990}. We use a spherical Navarro-Frenk-White (NFW) profile for the dark matter halo \citep{navarro96} implemented in \texttt{gala} \citep{gala2017,gala2020}. 
Throughout this work, the halo mass concentration, $c$, is set by the mass-concentration relation in equation 5 in  \citet{Munoz2011}. We first explore the Dw3 stream's evolution in a fixed potential (see black solid line in Figure \ref{fig:potential}) motivated by the enclosed mass estimates from GC kinematics \citep{Woodley2010}: $M(R<40~{\rm kpc}) = 9.7 \pm 3.3 \times10^{11}$ M$_{\odot}$ (see black data point in Figure \ref{fig:potential}). In this potential, M$_{200}$ = $9.4 \times10^{12}$ M$_{\odot}$ and $c = 8.05182$. We refer to this potential as the fiducial potential. To explore the minimum halo mass of Cen A in which we can reproduce the Dw3 stream, we also vary the potential parameters of Cen A in a mass range of 0.1 - 1.9 $\times$ M$_{200,\rm fiducial}$ with $c$ updated accordingly from \citet{Munoz2011} (see the ten colored lines in Figure \ref{fig:potential} and the potential values in Table \ref{tab:pot}). 
We also plot the projected separation of Dw3 and Cen A (79 kpc) as a green vertical line for reference. 

\begin{deluxetable}{lcccc}
\tablecaption{Summary of potentials}
\tablenum{2}\label{tab:pot}
\tablewidth{0pt}
\tablehead{\colhead{} &
\colhead{M$_{200}$} &
\colhead{$c$} &
\colhead{$M(R<40$ kpc)} &
\colhead{$-{\rm ln}\mathcal{L}$\tablenotemark{a}} \\
\colhead{} &
\colhead{M$_{\odot} \times 10^{12}$}  &
\colhead{} &
\colhead{M$_{\odot} \times 10^{11}$} &
\colhead{} 
}
\startdata
{\bf fiducial} &  {\bf 9.4}& {\bf 8.0518}& {\bf 9.7} & 3.16\\
Halo 1 & 0.94   & 10.0669 & 3.43 & -\\
Halo 2 & 2.82   & 9.0493 &  5.73 & 29.85\\
Halo 3  & 4.70  & 8.6118 &  7.21 & 7.83 \\
Halo 4  & 6.58  & 8.3353 &  8.35 & 4.84\\
Halo 5  & 8.46  & 8.1345 &  9.29 & 3.50\\
Halo 6  & 10.34 & 7.9777 &  10.1 & 2.86\\
Halo 7  & 12.22 & 7.8495 &  10.8 & 2.80\\
Halo 8  & 14.10 & 7.7413 &  11.4 & 2.22\\
Halo 9  & 15.98 & 7.6479 &  12.0 & 1.89\\
Halo 10 & 17.86 & 7.5658 &  12.5 & 2.09
\enddata
\tablenotetext{a}{See discussion of these values in Section \ref{sec:varycena}.}
\end{deluxetable}

\subsection{Control points \& coordinate system}\label{sec:control}
To explore which dark matter halo potentials of Cen A allow us to reproduce the straight stream emerging from Dw3, we need to compare stream models on various orbits to the data. For this purpose, we place 13 control points, $x_n$, by eye, along the center of the stream (see orange points in Figure \ref{fig:cena_data}) that we treat as our data points for delineating the track of the stream on the sky. 

\begin{figure}
\centerline{\includegraphics[width=\columnwidth]{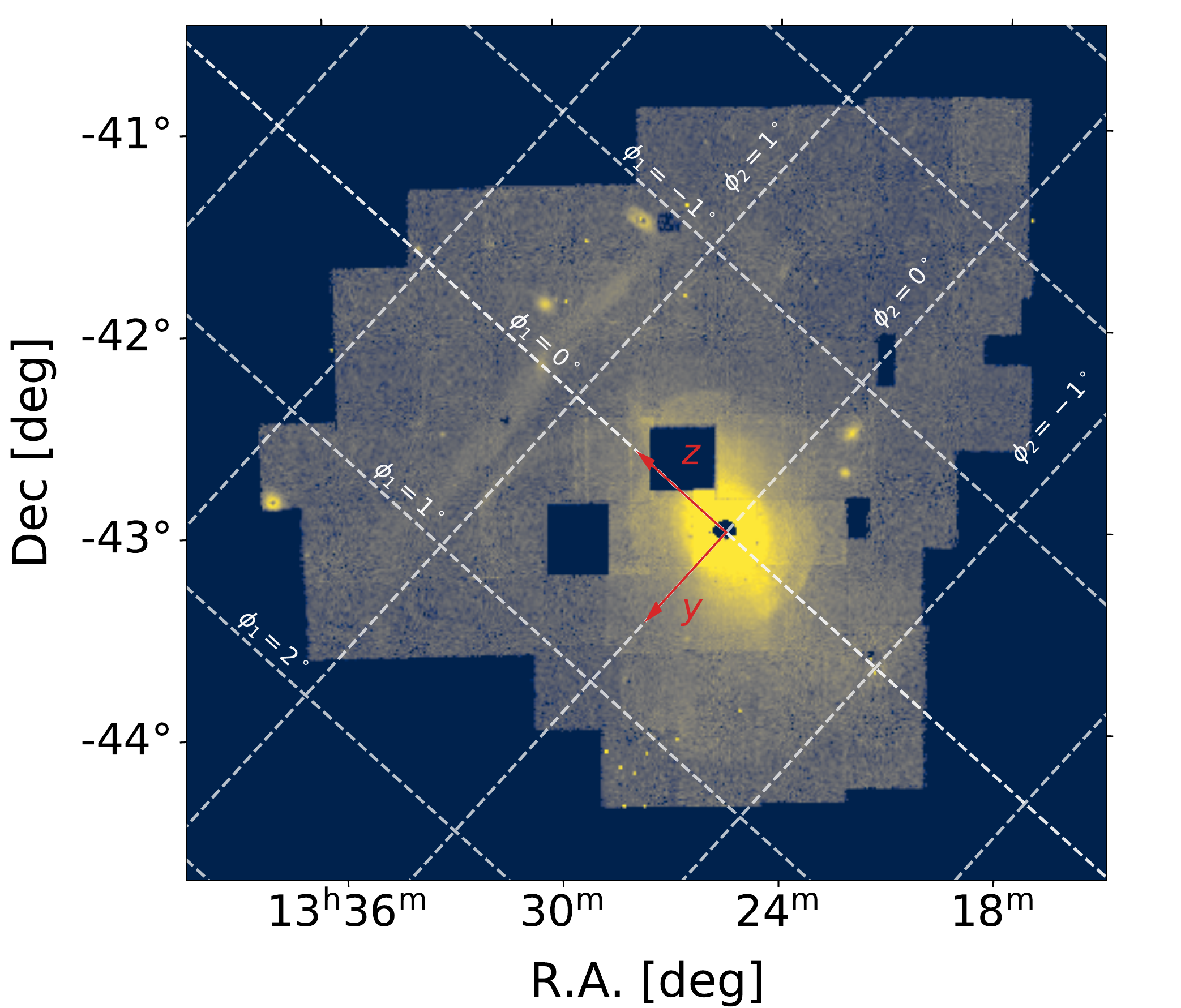}}
\caption{ The same data as shown in Figure \ref{fig:cena_data}, but here we overplot the two other coordinate system used in the paper: the galactic coordinate system ($x,y,z$) used for stream modeling (red), as well as the $\phi_1, \phi_2$ coordinate system which is centered on Cen A and rotated by  47.76 $\deg$, such that Dw3 is located at $\phi_1 = 0$ (white). We use the latter to visualize our mock stream fits throughout the paper. } 
\label{fig:coords}
\end{figure}

We also define two new coordinate systems which we summarize in Figure \ref{fig:coords}.
The first new coordinate system has  longitude and latitude as $\phi_1, \phi_2$ respectively and is centered on the Cen A stellar distribution and rotated by  $\alpha = 47.76^\circ$  to roughly align the Dw3 stream with constant latitude. We use the ICRS coordinates of Dw3 and Cen A respectively to define this rotated system such that Cen A is at $(\phi_1, \phi_2) = (0, 0)^\circ$ and Dw3 is at $\phi_1=0^\circ$ (see white dashed lines in Figure \ref{fig:coords}).\footnote{In detail, we define this coordinate frame using the \texttt{Astropy} coordinate transformation system \citep{astropy13,astropy18}, making use of the \texttt{SkyOffsetFrame}.}
In this coordinate system, $\phi_{1}$ is the longitude along the stream (and increases in the direction of motion), and $\phi_{2}$ indicates the direction perpendicular to the stream.
For the control points, we define error bars of $\sigma_n = 0.01$ degrees in the direction of $\phi_{2}$ for each data point, $x_n$, in this frame. This error is set by how far we can move each control point while the point still visually appears to be at the stream center. While a change in $\sigma_n$ can lead to a different exact value of a log-likelihood function, it will not change our assessment of the best fit.

We then define a coordinate system centered on the three-dimensional position of Cen A by translating to the distance to Cen A (3.8 Mpc) such that the $x$ direction of the new Cartesian coordinate system points in the direction to the Milky Way's center. In this coordinate system, the projected distance between Cen A and Dw3 is fixed to 79 kpc. The other two directions  (for $y$ and $z$) are shown in Figure \ref{fig:coords} (see red arrows).
We use this coordinate system to transform stream models generated in the Cen A halo back to the $(\phi_{1}, \phi_{2})$, observed sky coordinates in order to evaluate the fits (see Section \ref{sec:streamfit}). 
Lastly, we transform the simulated streams to the ICRS coordinates to record the observable radial velocities and distance gradients along the model streams. 

\subsection{Stream modeling}\label{sec:streamgenerator}
To simulate Dw3's stream evolution in Cen A's halo, we generate model streams using the ``particle spray''-method \citep{Fardal2015} implemented in \texttt{gala} \citep{gala2017,gala2020}. 
For a given position and velocity of Dw3, we integrate its orbit backwards in time over 6 Gyr.
From the orbit end point, we integrate the orbit forwards while we release two stars from each Lagrange point per Myr, with a spread in position and velocity dispersion set by the progenitor's mass \citep[see][]{Fardal2015}. We include self-gravity of the progenitor, which we model as a Plummer sphere \citep{plummer1911} with $m_{\rm Dw3} = 10^{8}$ M$_{\odot}$ and $b = 2.5$ kpc to ensure a realistic length of the stream \citep{gibbons14}. This progenitor mass produces model streams with similar widths to the observed stream.
The widths of the streams are left as free parameters in our assessments of the fits to the control data. 
We do not update the mass of the progenitor throughout its orbit as this will only slightly affect the dispersion in position and velocity of the released particles (but see how we consider the evolving mass of the progenitor below).

Using the \citet{Fardal2015} mock stream generator enables us to rapidly explore many different Dw3 orbits as compared to full N-body simulations. One major caveat with this method, however, is that, in reality, Dw3 will not strip stars if the tidal radius is larger than the actual radius of Dw3. This is not taken into account when the ``particle spray" stream code releases stars in each time step. 
To ensure that we do not include streams that would not have formed (i.e. if the progenitor is on first infall or orbiting too far from Cen A's center to strip), we also calculate the Jacobi radius, which can be expressed as:
\begin{equation}\label{eq:jacobi}
    r_{J} = R\left(\frac{m_{\rm prog}}{M(<R)}\right)^{1/3}
\end{equation}
where $R$ is the galactocentric radius at a given time, $m_{\rm prog}$ is the Dw3 progenitor mass, and $M(<R)$ is the enclosed mass of the potential at the current radius of the orbit. The Jacobi radius changes throughout the evolution of a non-circular orbit and with mass loss of the progenitor. From \citet{denja:2016} figure 9, we know that Dw3's present day tidal radius is $r_{\rm tidal, Dw3} \sim 3.3$ kpc. Therefore, we add an additional post-processing step to our mock stream generator, where we remove stars that were released with the ``particle spray'' method while the Jacobi radius for the simulated dwarf was $r_{J} > 2\times r_{\rm tidal, observed}$. We refer to this as the ``Jacobi radius criterion'' throughout the paper. While we do not update the mass of the progenitor in the stream modeling, we do assume that Dw3 decreases in mass from $1\times 10^8$ M$_{\odot}$ to $3 \times 10^7$ M$_{\odot}$ throughout its 6 Gyr of evolution when we calculate the Jacobi radius (see observational motivation for this in Section \ref{sec:data}). 
Note that a cut of $2\times r_{\rm tidal, observed}$ is quite conservative and will often lead to stars stripping along the entirety of Dw3's orbit.

\subsection{Comparison to observational data}\label{sec:streamfit}
To evaluate how well our model stream tracks fit the control data points, we compute the center point of the simulated streams, $y_n$, evaluated at the location of the control data, $x_n$, in the ($\phi_1,\phi_2$) coordinate frame in degrees, where the Dw3 stream is horizontal. This allows us to ignore any horizontal error bars. We compute the log-likelihood:

\begin{equation}
   {\rm ln}\mathcal{L} = -\frac{1}{2}\sum_n\frac{(x_n - y_n)^2 }{\sigma^2}\label{eq:L}
\end{equation}
where $x_n$ are the 13 control data points, $y_n$ are the 13 simulation evaluation points, and $\sigma = 0.01$ deg is the vertical error in the $\phi_2$-direction.
We pick a regularly spaced grid of longitude ($\phi_1$) values by eye and assume that we know these values perfectly. The density of the stream has some error in the perpendicular ($\phi_2$) direction, and we place the control points along the stream's center by eye with a precision, which we assume to be $\sigma = 0.01$ deg. 
The best fit is when the model stream center points, $y_n$, have the minimum squared deviation from the control data points, $x_n$ (see Equation \ref{eq:L}). 
Throughout this work, we evaluate  ${\rm ln}\mathcal{L}$ both before and after we apply our Jacobi radius criterion (see Section \ref{sec:streamgenerator}), and we discard model streams that do not cover the extent of the data points by setting ${\rm ln}\mathcal{L} = -$inf for those cases.
The length, width and surface density of the model streams are left as free parameters throughout this paper \citep[see][for an alternative approach including these parameters]{erkal2017}.

\subsection{$N$-body follow up}\label{sec:nbodyfollowup}
To check whether the \citet{Fardal2015} mock stream generator produces a similar stellar stream under more realistic stripping conditions, we also run follow up $N$-body simulations of our best fit stream models. In particular, we set up Dw3 as a two-component mass model. We again represent the stellar component of Dw3 as a Plummer sphere \citep{plummer1911} with a mass of $1\times10^8\ {\rm M}_{\odot}$, and scale radius $b=2.5$ kpc, but now also include a Hernquist sphere \citep{hernquist1990} to represent the dark matter ($r_s$ = 5 kpc, M$_{\rm dm} = 64 \times {\rm M}_* = 6.4 \times 10^9$ M$_{\odot}$\footnote{Note that abundance matching predicts higher dark matter to baryon ratios for dwarfs at this stellar mass scale \citep[e.g.,][]{Behroozi2010}, but since we are interested in the stripping of the stellar component, which happens after the outer parts of the dark matter is gone, we do not explore more massive Dw3 dark matter profiles in this work.}. 
We use the galactic dynamics python package \texttt{Agama} to sample $10^6$ particles from the Plummer stellar profile and 1.6 $\times ~10^6$ particles from the Hernquist dark matter profile. We then initialize this dwarf galaxy at the end point of the integration for the best fit model stream,  
and integrate the dwarf for 6 Gyr within the same fixed fiducial Cen A potential (see black line in Figure \ref{fig:potential}) using the $N$-body simulation code \texttt{GCD+} \citep{Kawata2003}.

\section{Results}\label{sec:results}
In this Section, we present the results of our Dw3 stream analyses.  We first fix the Cen A dark matter halo profile to explore which orbit solutions can reproduce the straight stream emerging from Dw3 (section \ref{sec:fixedcena}). We follow up our best fit stream with an N-body code (Section \ref{sec:nbody}). We then explore the degeneracy between the dark matter halo mass and stream morphology in Section \ref{sec:halodegeneracy}, and lastly search for the best-fit solution to the Dw3 stream control data while varying Cen A's dark matter potential in Section \ref{sec:varycena}. Throughout this Section, we assume a distance to Cen A of 3.8 Mpc and a distance to Dw3 of 3.88 Mpc (see Section \ref{sec:discussion} for the effects of varying Dw3's distance). We also assume an initial stellar mass for the progenitor of m$_{\rm Dw3}=  10^8$ M$_{\odot}$ and scale radius $b = 2.5$ kpc, when we simulate the stream. 
We summarize the simulation parameters used in this Section in Table \ref{tab:sim}.

\begin{deluxetable}{lcc}
\tablecaption{Summary of simulation set up}
\tablenum{3}\label{tab:sim}
\tablewidth{0pt}
\tablehead{\colhead{Parameter} &
\colhead{value} &
\colhead{unit}
}
\startdata
NFW halo Cen A\\
\hline
$M_{\rm 200, fiducial}$ &  $9.4\times 10^{12}$ & M$_{\odot}$\\
$c_{\rm 200, fiducial}$ & 8.05182 &\\
\hline
Hernquist sphere Cen A\\
\hline
$m_{\rm *, fiducial}$ &$1\times 10^{11}$ &M$_{\odot}$ \\
$a_{\rm *, fiducial}$\tablenotemark{a} &4.896& kpc \\
\hline
Plummer sphere Dw3\\
\hline
$m_{\rm Dw3}$  & $1\times10^8$ & M$_{\odot}$\\
$b_{\rm Dw3}$ & 2.5 & kpc\\
\hline
Other parameters\\
\hline
$d_{\rm CenA}$ &3.80& Mpc\\
$d_{\rm Dw3}$ &3.88& Mpc\\
$t_{\rm integration}$  & 6 & Gyr \\
$v_{\phi_1}, v_{\phi_2}$ & $-800$ to $800$ &km s$^{-1}$\\
$v_{\rm rad}^b$  & $-181.4$ & km s$^{-1}$  \\
$r_J$-cut & 2 $\times$~ 3.3 & kpc
\enddata
\tablenotetext{a}{From \citet{hernquist1990}. $^b$: relative between Cen A and Dw3. }
\end{deluxetable} 

\subsection{Dw3 stream morphology and orbit in a fixed Cen A dark matter halo}\label{sec:fixedcena}
\begin{figure*}
\centerline{\includegraphics[width=\textwidth]{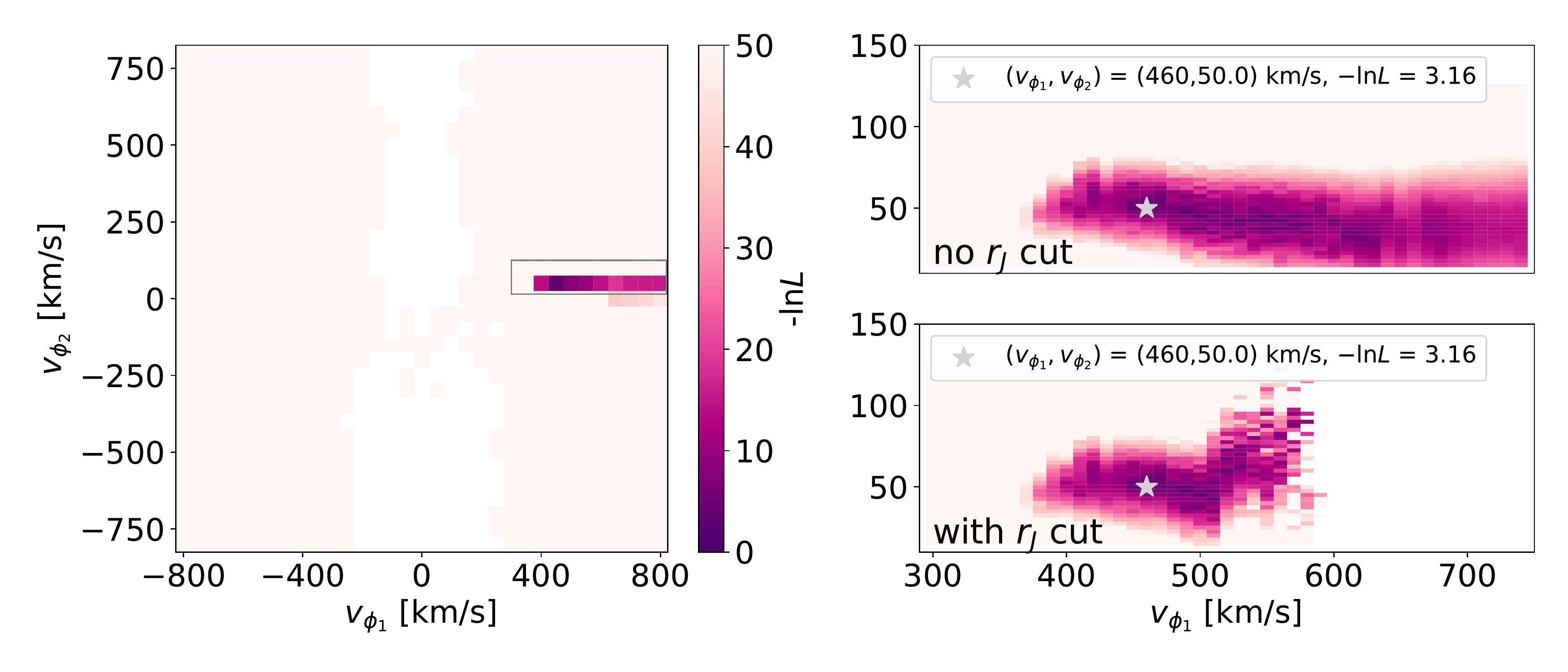}}
\caption{Summary of $-{\rm ln}\mathcal{L}$-values (Eq. \ref{eq:L}, see colorbar) for model streams initiated with a fixed radial velocity of $v_{rad} = -181.4$ km s$^{-1}$, but with various combinations of $v_{\phi_1}$  and $v_{\phi_2}$. 
The fits of each stream to the control data is evaluated as described in Section \ref{sec:control}. 
The blank space represents the region of the orbit parameter space, where the simulated streams were disregarded, since they were shorter than the extent of the control data.
Note that the maximal color bar value is $-{\rm ln}\mathcal{L}$ = 50 for visualization purposes, but that some of the fits are extremely poor with $-{\rm ln}\mathcal{L} > 1000$.
Left: There is a small pocket of orbits ($v_{\phi_1} 
\sim 300$--$800$ km s$^{-1}$ and $v_{\phi_2} \sim 40$--$100$ km s$^{-1}$) that produce good fits to the data (see gray rectangle).
Upper right: Zoomed version of gray rectangle in the left panel, where we resolve the combination of $v_{\phi_1}$ and $v_{\phi_2}$ which produced good fits to the data before any Jacobi criterion is applied.  
Lower right: Same as above, but after the Jacobi criterion has been applied. The model stream that produced the best fit has $v_{\phi_1}$ = 460 km s$^{-1}$ and $v_{\phi_2}$ = 50 km s$^{-1}$ with $-{\rm ln}\mathcal{L} = 3.16$ in both cases.}
\label{fig:orbits}
\end{figure*}

\begin{figure*}
\centerline{\includegraphics[width=\textwidth]{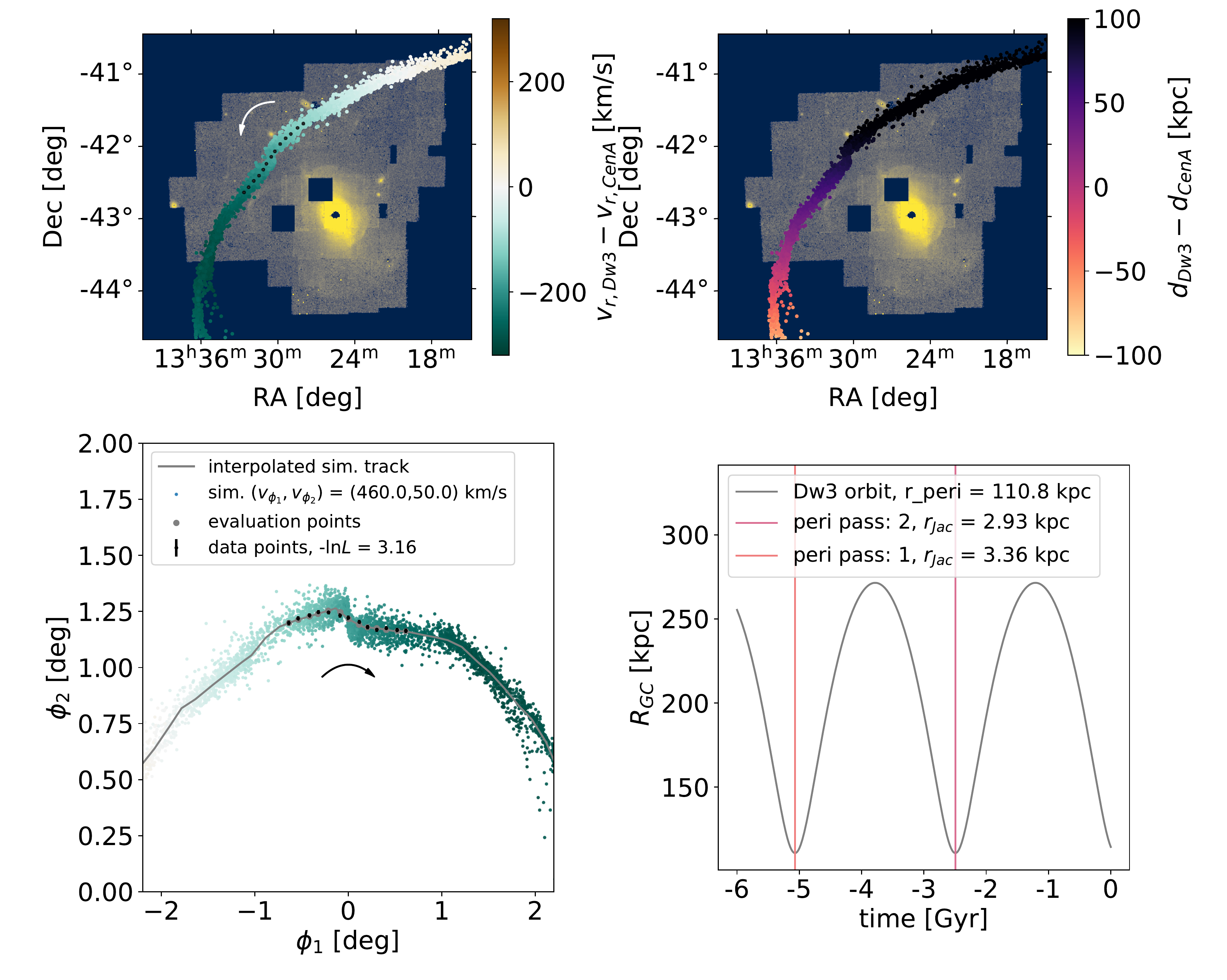}}
\caption{Visualization of the best fit model stream from Figure \ref{fig:orbits} with $v_{\phi_1}$ = 460 km s$^{-1}$ and $v_{\phi_2}$ = 50 km s$^{-1}$ and $-{\rm ln}\mathcal{L} = 3.16$.
Upper panels: The model stream transformed to ICRS coordinates, colored by radial velocity in Cen A's rest frame (left) and distance along the stream with respect to Cen A (right), overplotted on the Cen A image data. The white arrow indicates the direction of motion of Dw3, and the black points are the 13 control data points, $x_n$ (see also orange points in Figure \ref{fig:cena_data}). 
Lower left: The best fit model stream shown in ($\phi_1,\phi_2$)-space in degrees (see Section \ref{sec:control}). The orange points are the 13 control data points, $x_n$, the green points (colored by radial velocity) are the simulated model stream, the gray line is the center point of the simulated model stream, the gray dots are the 13 simulated stream center points, $y_n$ at the $\phi_1$-location of the control data. The black arrow indicates the direction of motion of Dw3.
Lower right: The evolution of the 3D position in kpc of the progenitor with respect to Cen A's center as a function of time (black line, t = 0 at present day). The vertical lines mark the pericenter passages of the progenitor. 
}
\label{fig:fit_fixedhalo}
\end{figure*}

To explore the Dw3 stream evolved in the fiducial potential of Cen A (see Figure \ref{fig:potential}, solid black line), we  first fix Dw3's position and the relative radial velocity between Dw3 and CenA (see Table \ref{tab:cena}). 
Thus, the only free parameters in our initial search for orbits that can reproduce the Dw3 stream are the two other Dw3 velocity components. We evaluate the fits to the data in the rotated and shifted coordinate system where Cen A is located at $(\phi_1, \phi_2) = (0,0)$ deg, with $v_{\rm rad} = 0$ km s$^{-1}$. The $\phi_1$-direction is parallel to the stream, and the $\phi_2$-direction is perpendicular to the stream (see Section \ref{sec:control}).

We set up a grid of velocities with $v_{\phi_1}$ and $v_{\phi_2}$ ranging from $-800$ to 800 km s$^{-1}$ in steps of 50 km s$^{-1}$. For each velocity combination on this grid, we initialize a model stream simulation (see Section \ref{sec:streamgenerator}), integrate the progenitor for 6 Gyr in steps of 1 Myr (see discussion of integration time in Section \ref{sec:deg}) and evaluate the fit of the simulated stream to the actual data as explained in Section \ref{sec:streamfit}. For each model stream and for each velocity combination on the grid, we store the value of $-{\rm ln}\mathcal{L}$. 

In Figure \ref{fig:orbits} (left), we clearly see that there is one pocket of orbital solutions that can generate a good fit to the data (see gray box). These solutions all reside in the part of velocity space where the progenitor is moving rapidly parallel to the stream ($v_{\phi_1}$) and slowly perpendicular to the stream  ($v_{\phi_2}$). 
We explore all directions of motion of Dw3, but the model streams clearly match the data best for orbits moving in the positive $v_{\phi_1}$-direction. This is due to the ``S''-shape of the Dw3 stream with the southern (leading) part of the stream slightly closer to Cen A's center than the northern (trailing) part. We discuss the ``S''-shape further later in this section.
Note that empty parts of the velocity grid presented in Figure \ref{fig:orbits} (left), represent failed model streams that did not cover the extent of the control data points during the 6 Gyr of evolution. 
While we disregard model streams that are shorter than the extent of the control data, the maximal extent of the model stream lengths is a free parameter in our fits (see Section \ref{sec:streamfit}). Therefore, the model streams are often longer than the actual control data. The exact length of streams is set by a  combination of integration time and the complex escape conditions from the progenitor, which can also include dark matter. The debris at the end of streams will often be of lower surface density and might not be observable. See Section \ref{sec:nbody} for a discussion of progenitor escape conditions. 

To ensure that we resolve the best fit solution, we initiate a new, zoomed, velocity grid where $v_{\phi_1}$ ranges from 300--750 km s$^{-1}$ in steps of 10 km s$^{-1}$ and $v_{\phi_2}$ ranges from $15$--125 km s$^{-1}$ in steps of 2.5 km s$^{-1}$. We initialize new model streams with these new combinations of orbital parameters, and use the same setup as before. We present the results for these model streams in the upper right panel of Figure \ref{fig:orbits}, where we have included all stars in our assessment of $-{\rm ln}\mathcal{L}$. 
Note that the $-{\rm ln}\mathcal{L}$-space is not convex. Thus, classical optimization methods will be non-trivial in this space, which is why we do a brute-force grid search for the best solution. 
The gray star indicates the orbit that produced the best fit ($-{\rm ln}\mathcal{L}$ = 3.16 and ($v_{\phi_1}, v_{\phi_2}$) = (460, 50) km s$^{-1}$). 
Some of these velocity combinations place Dw3 on orbits which would not lead to tidal stripping, however.

In the lower right panel of Figure \ref{fig:orbits}, we therefore evaluate $-{\rm ln}\mathcal{L}$ after we apply the Jacobi criterion (see Eq. \ref{eq:jacobi} and  Section \ref{sec:streamgenerator}) to ensure more realistic stellar stripping conditions. Note that for many of the high $v_{\phi_1}$ orbits ($> 590$ km s$^{-1}$), most stars are removed with this cut, and the model streams do not cover the extent of the data (see blank space). We find that the best fit solution is the same as before the Jacobi cut. The Jacobi radius for the progenitor on this orbit remains smaller than 2 times the observed tidal radius for the duration of the 6 Gyr. Hence, no stars are removed in the Jacobi radius post-processing step. 
There are several orbits in this fiducial potential that produce good fits (see dark colors), and the Dw3 stream is easily reproduced for high $v_{\phi_1}$-values and low $v_{\phi_2}$-values.
We summarize the parameters of our best fit in the fiducial halo in Table \ref{tab:bestfit}.

In Figure \ref{fig:fit_fixedhalo}, we visualize this best fit model stream from Figure \ref{fig:orbits}. We show the model stream overplotted on the actual image data of Cen A in ICRS coordinates colored by the radial velocity of the stream in Cen A's rest frame (upper left) and the distance gradient along the stream with respect to Cen A (upper right). Most of the stream is moving towards us  (with radial velocities relative to Cen A ranging from $v_{\rm rad} = -116.0$ for the most northern control data point to $-244.5$ km s$^{-1}$ for the most southern control data point). The distance gradient along the stream ranges from 100.6 kpc with respect to Cen A for the most southern control data point to 55.4 kpc for the most northern control data point.
We also visualize the best fit in  $({\phi_1},{\phi_2})$-space (lower left) where the stream is moving towards positive ${\phi_1}$ (see arrow).  
Note here the pronounced ``S''-shape of both the model stream and the control data, which helped constrain the direction of motion (i.e., the leading arm is the part of the stream that is closer to Cen A's center).
We show the evolution of the galactocentric radius in time for the progenitor's orbit in the lower right panel (t = 0 at present day). 

For this fit, Dw3 has completed two pericenter passes in 6 Gyr and is approaching pericenter at present day (see vertical lines in the lower right panel). Note the difference between the Jacobi radius at the first ($r_J = 3.42$ kpc) vs second ($r_J = 2.97$ kpc) pericenter. This is due to the mass loss of the progenitor included in our $r_J$ calculation (see Section \ref{sec:streamgenerator}). 
Interestingly, for most of the best-fit solutions, Dw3 is close to pericenter at present day. The progenitor moves most rapidly at pericenter. The fact that the best fit stream models are mostly at pericenter at present day is likely because the stream needs to move fast in the $\phi_1$-direction to be so straight, while remaining bound to Cen A.  Note also that the observed blueshifted line-of-sight velocity of Dw3 with respect to Cen A \citep[$v_r = -181.4$ km/s reported by][]{dumont2022} is consistent with the co-rotating line-of-sight velocity trend found by \citet{mueller2021} for 21 out of 28 of Cen  A's satellites with measured velocities. The proper motion of the best fit stream orbit in Figure \ref{fig:fit_fixedhalo}, however, ($\mu_{\alpha \rm  cos(\delta)}$, $\mu_\delta$) = ($0.0191$,$-0.0164$) mas/yr, moves Dw3 in a direction out of Cen A's planar satellite structure \citep{mueller2018}.

\begin{deluxetable}{lcc}
\tablecaption{Summary of the best fit in the fiducial halo}
\tablenum{4}\label{tab:bestfit}
\tablewidth{0pt}
\tablehead{\colhead{Parameter} &
\colhead{value} &
\colhead{unit}
}
\startdata
($v_{\phi_1}$,$v_{\phi_2}$,$v_r$)&(460,50,$-181.4$) &[km s$^{-1}$]\\
 $-{\rm ln}\mathcal{L}$ &3.16 &\\
 number of peri passes  & 2&\\
 $v_{\rm rad}$-range\tablenotemark{a}   &  $-116.0$ -- $-244.5$  & [km s$^{-1}$]\\
 distance range\tablenotemark{a}  & 100.6 -- 55.4 &[kpc]
\enddata
\tablenotetext{a}{Listed with respect to Cen A's rest frame from the most northern control data point to the most southern control data point.} 
\end{deluxetable}

\subsection{Verifying the best fit with an $N$-body code}\label{sec:nbody}
\begin{figure*}
\centerline{\includegraphics[width=\textwidth]{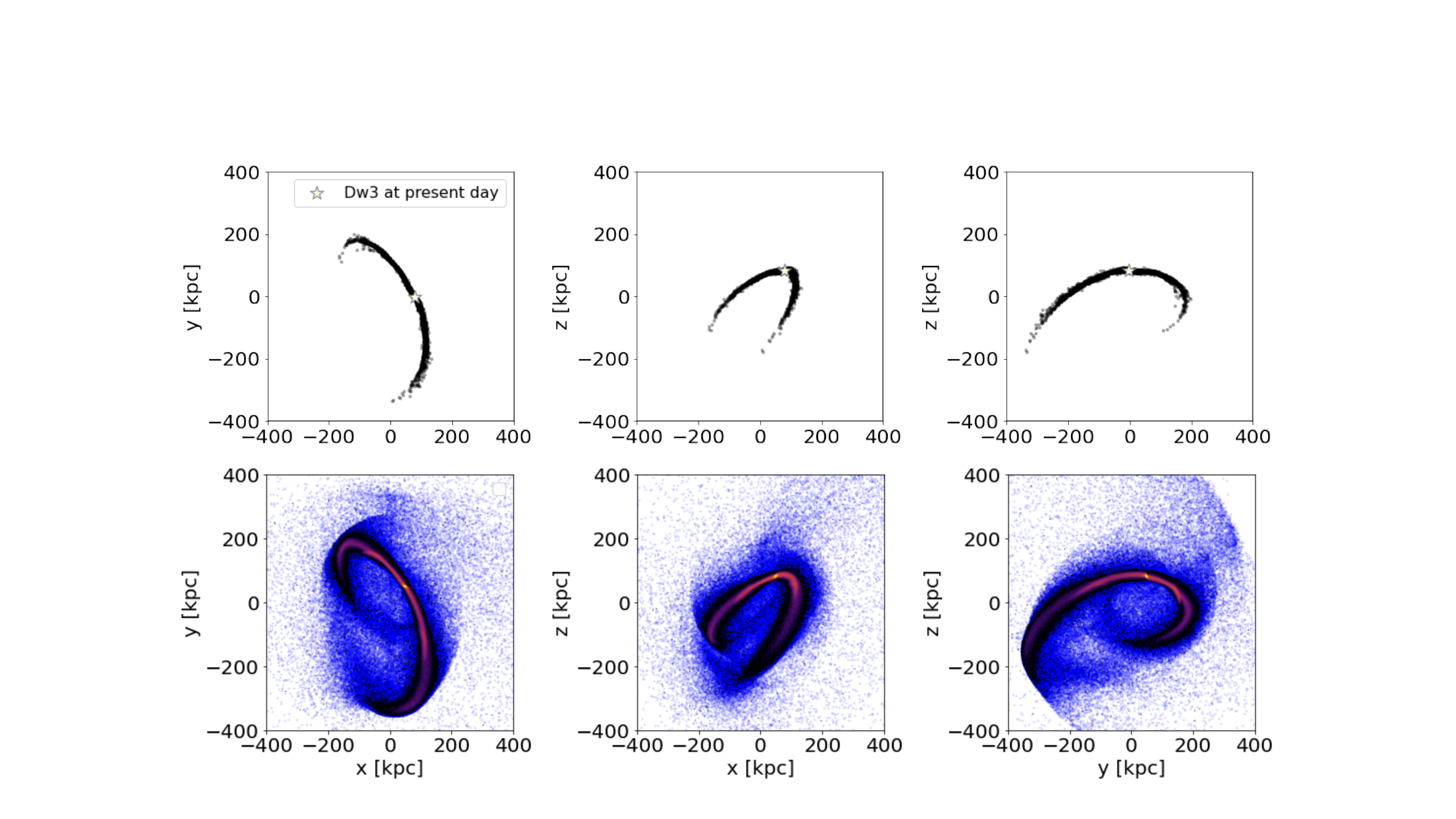}}
\caption{Comparison between the best fit model stream evolved with the \citet{Fardal2015} mock stream generator (top) and the \texttt{GCD+} \citep{Kawata2003} N-body follow-up (bottom). Both runs are shown in galactocentric coordinates centered on Cen A at ($x,y,z$) = (0,0,0), and Dw3 was evolved for 6 Gyr in the fiducial potential in both cases. 
Top panels: Black particles show the stars in the model stream, and the white star marks the progenitor position at present day.
Bottom panels: Blue particles show the extended dark matter, and the inferno color map shows the stars. The Dw3 remnant is visible as an overdensity of stars. 
The stellar component of both the model stream (top) and the N-body stream (bottom)  look very similar.}
\label{fig:nbody}
\end{figure*}
We explore whether the best fit Dw3 stream can be reproduced under more realistic stellar stripping conditions with the \texttt{GCD+} \citep{Kawata2003} N-body code (see Section \ref{sec:nbodyfollowup} for the detailed setup of the simulation).
We initialized the simulation from the end point of the 6 Gyr orbit for the best fit model stream in the fiducial halo, where $(x,y, z)=(-128.999, 170.042,-43.190)$ kpc and $(v_x,v_y,v_z)=(-3.378,-279.266,-158.709)$ km s$^{-1}$, and integrate Dw3 forwards in the fiducial potential for 6 Gyr to its present day position. 

In Figure \ref{fig:nbody}, we show the results of the comparison between the model stream from Figure \ref{fig:fit_fixedhalo} in Cen A galactocentric coordinates (top) and for the \texttt{GCD+} \citep{Kawata2003} N-body run in galactocentric coordinates (bottom) at present day.  
We find that the dark matter (blue particles) extends beyond the stellar component (inferno color map), as expected. 
The morphology of the stellar streams produced in both the
\citet{Fardal2015} model stream case (top) and the N-body run (bottom) are very similar. The Dw3 position at present day is marked with a white star in the top panel, and is visible as an overdensity of stars in the bottom panel. 
For this particular run, we find that there are $1.4 \times 10^7$ M$_{\odot}$ stars and $7.4 \times 10^7$ M$_{\odot}$ dark matter remaining within 5 kpc of the Dw3 remnant in the N-body run at present day. This stellar mass is lower than the present day observed mass of the remnant, but these numbers depend on the detailed setup of Dw3 and could be fine-tuned to better match to the data (see Section \ref{sec:nbodyfollowup}).
While the presence of dark matter in the Dw3 remnant changes the exact escape conditions, which are also more simplistic in the model stream case (see Section \ref{sec:streamgenerator}), we conclude that the \citet{Fardal2015} model streams produce good approximations to the Dw3 stellar stream evolution.

\subsection{Halo mass vs stream morphology degeneracy}\label{sec:halodegeneracy}
\begin{figure*}
\centerline{\includegraphics[width=0.95\textwidth]{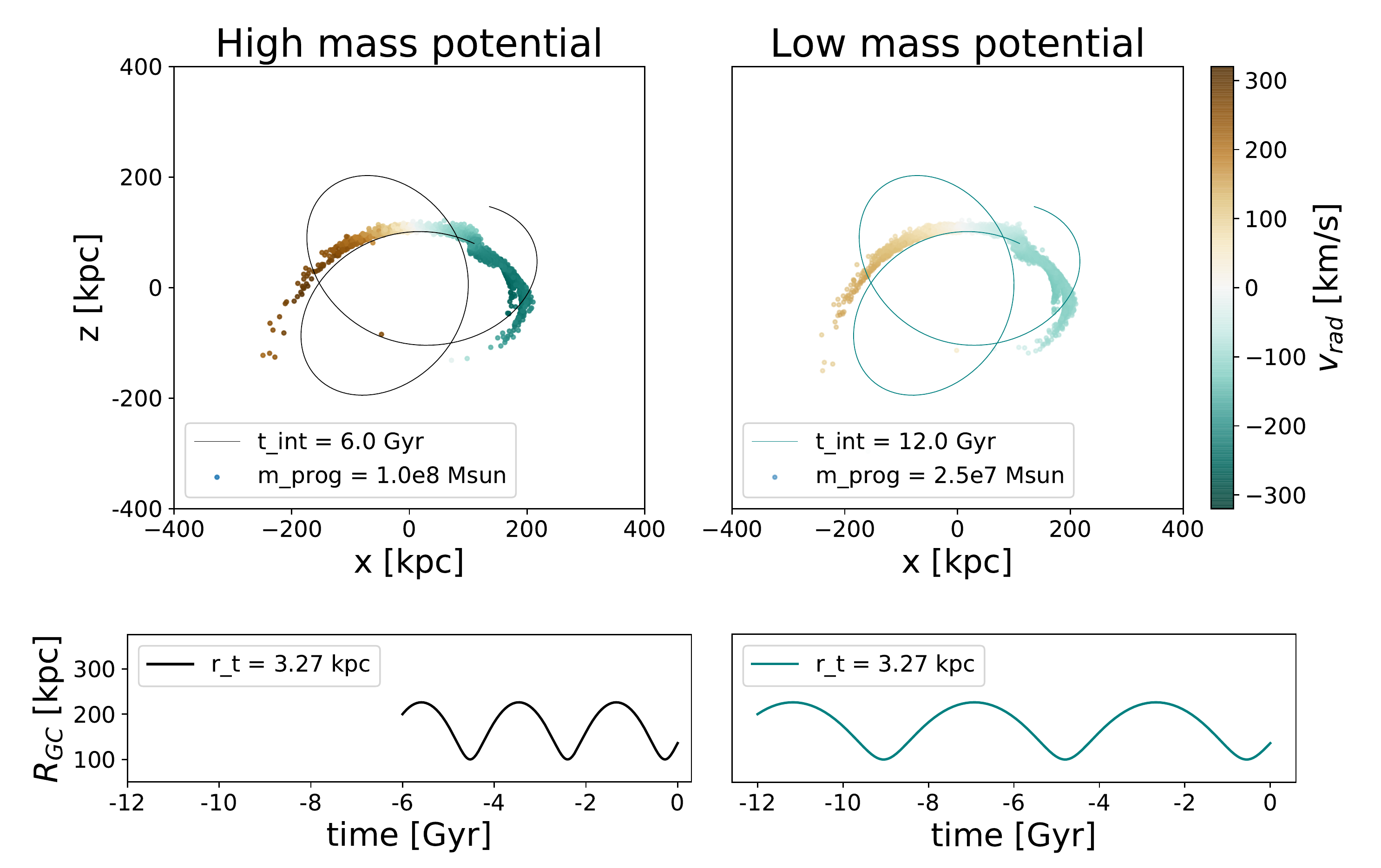}}
\caption{
Comparison between two streams evolved in a high mass (left) and low mass (right) halo. 
Top Left: model stream evolved for 6 Gyr in fiducial halo (colored by radial velocity in Cen A's rest frame) and progenitor orbit (black line) projected in $x$ and $z$ galactic coordinates in kpc, where Cen A is located at ($x,z$)=(0,0). The progenitor has an  initial mass of $10^8$ M$_{\odot}$.  
Top right: Model stream evolved for 12 Gyr in a halo mass scaled down by a factor of four as compared to the fiducial halo. The teal line shows the orbit of the progenitor. The progenitor mass in the right panel is also scaled down by a factor of four, and its orbital velocity is scaled down by a factor of two. 
The two streams have identical morphologies but very different radial velocities. 
Bottom left: The model stream's 3D galactocentric distance with respect to Cen A's center as a function of time.
Bottom right: Same as bottom left, but now for the progenitor with a fourth of the mass and half of the orbital velocity. Note how the tidal radius, pericenter and apocenter distances are identical in the two cases.
}
\label{fig:halomassdegeneracy}
\end{figure*}

Our goal is to test whether the Dw3 stream can provide limits on the dark matter halo mass distribution of Cen A. But before we start, it is important to understand the degeneracies that exist between dark matter halo mass and the properties of the progenitor's orbit, mass, and resulting stream morphology. 

In Figure \ref{fig:halomassdegeneracy}, we simulate a stream with an orbit similar to the best fit stream in Figure \ref{fig:fit_fixedhalo}, now in the galactic coordinate system of Cen A (see Section \ref{sec:control}) in both the fiducial dark matter halo (see upper left panel), as well as in a NFW dark matter halo with a fourth of the mass, but for a fixed scale radius. 
In the case of Dw3 in Cen A's halo, we know the projected distance between Cen A and the dwarf. We have therefore initiated the two streams with the same physical separation from Cen A's center. 
The difference between the high mass (left) and low mass (right) progenitor orbits, is that the right stream, evolved in the lower mass halo, has a velocity scaled down by a factor of two compared to the left. 
The enclosed mass difference at pericenter for the left and right scenario is a factor of four smaller for the low mass halo (right). In this example, we therefore scaled down the Dw3 progenitor mass by a factor of four in the low mass halo (right) and integrate for twice as long \citep[see][for an analytic expression of these scalings in a logarithmic potential]{Johnston2001}. 

We find that the morphologies of the two streams are identical in the high mass vs low mass case in Figure \ref{fig:halomassdegeneracy} (except for scatter induced by the stream modeling technique). The apocenter and pericenter distance with respect to Cen A's center (lower panels), and the tidal radius at present day are also identical (in this toy example, we do not assume any mass loss of the progenitor). However, the radial velocity of the two streams are very different (see color bar). While the stream evolved in the fiducial halo (left) has a radial velocity of $-181.4$ km s$^{-1}$, Dw3, evolved in the lower mass halo, needs to have lower radial velocity by a factor of two ($-90.7$ km s$^{-1}$) in order to reproduce the same morphology. 
Thus, knowing just the one radial velocity of Dw3, helps break the halo mass vs stream morphology degeneracy. Note also how the radial velocities along the streams are very different in the two scenarios. 

\subsection{Varying the halo mass}\label{sec:varycena}
To test the constraining power of the Dw3 stream on Cen A's halo mass, we explore whether the Dw3 stream can be reproduced in the ten different potentials shown in Figure \ref{fig:potential} (see also Table \ref{tab:pot}). We test if there is a lower limit on the halo mass in which the stream morphology can not be reproduced, given the observational constraint of the radial velocity of Dw3 ($v_{\rm rad} = -181.4$ km s$^{-1}$; \citealt{dumont2022}). As we do not expect the  nature of the orbital solution to change due to a change in halo mass (see Figure \ref{fig:halomassdegeneracy}), we use the same velocity resolution in $v_{\phi_1}, v_{\phi_2}$ (i.e., stepping 10 km s$^{-1}$ in $v_{\phi_1}$ and 2.5 km s$^{-1}$ in $v_{\phi_2}$) as presented in Figure \ref{fig:orbits}, but spanning a larger range in velocities to capture the differences in each halo. For each velocity combination, we generate a stream, and record the $-{\rm ln}\mathcal{L}$ value, but now in ten different halos. 
In this Section, we only present $-{\rm ln}\mathcal{L}$ values and model streams after we apply the Jacobi cut (see Section \ref{sec:streamgenerator}) to ensure physical tidal stripping conditions. 
We again fix the distances to Cen A, the distance to Dw3, the radial velocity and mass profile of Dw3. 

\begin{figure*}
\centerline{\includegraphics[width=0.85\textwidth]{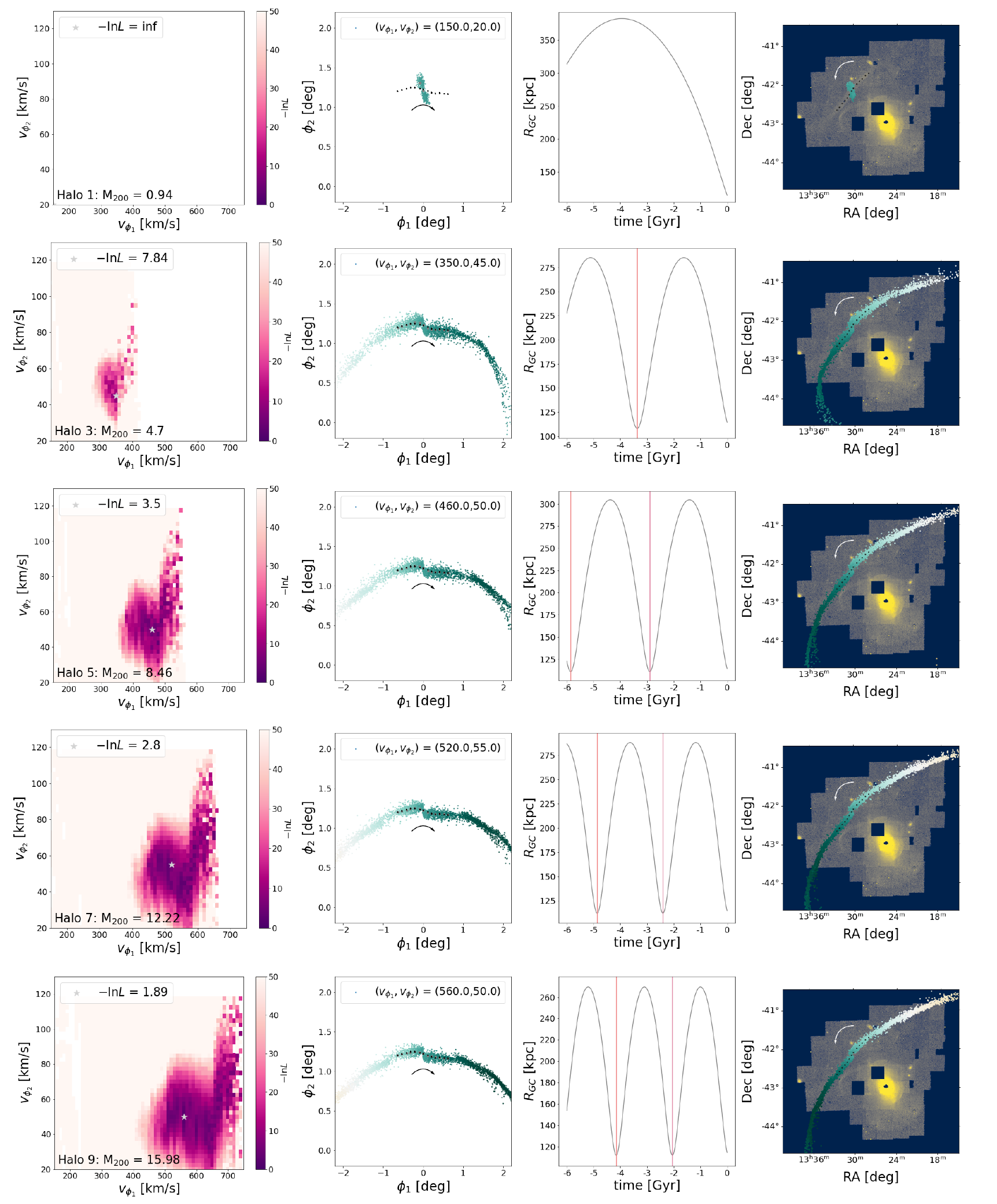}}
\caption{Summary of best fit model streams in various halos. {\bf Left panels:}  $-{\rm ln}\mathcal{L}$ for all model streams initiated with $v_{\phi_1}$ = 150-750 km s$^{-1}$ in increments of 10 km s$^{-1}$ and $v_{\phi_2}$ = $20$ - 120 km s$^{-1}$ in increments of 2.5 km s$^{-1}$ evolved for 6 Gyr in five different dark matter halos. The Jacobi radius cut has been applied in all panels (see Section \ref{sec:streamgenerator}). 
All M$_{200}$ values are listed in units of $10^{12}$ M$_{\odot}$.
The gray stars mark the combination of orbital parameters ($v_{\phi_1}$,$v_{\phi_2}$) that produced the best fit model stream in each halo.
Dark colors indicate pockets of best-fit solutions (see colorbar in Figure \ref{fig:orbits}). 
Blank spaces in the left parts of the panels indicate orbits for which the model stream did not cover the extent of the data after the Jacobi cut. 
{\bf 2nd panels:} Visualization of the best fit model stream in ($\phi_1,\phi_2$)-space in each halo after the Jacobi radius cut, colored by the radial velocity along the stream in Cen A's rest frame (see colorbar in Figure \ref{fig:fit_fixedhalo}). 
For Halo 1 (top), the model stream does not cover the extent of the data for any orbit after the Jacobi cut, so we show an example of an orbit with ($v_{\phi_1}$,$v_{\phi_2}$) = (150,20) km s$^{-1}$.
The black points show the 13 control data points ($x_n$), and the gray points, ($y_n$), show the model stream center points at the $\phi_1$ location of the control data (note that these are hard to see when the control data points overlap well with the center points of the simulated stream). The black arrow indicates the direction of motion of each model stream. 
{\bf 3rd panels:} The 3D position of Dw3 with respect to Cen A's center as a function of time. 
{\bf Right panels:} The best fit model streams colored by radial velocity in Cen A's rest frame and presented in ICRS coordinates, overplotted on the Cen A image data.  The white arrow indicates the direction of motion for each model stream. 
}
\label{fig:fit_grid}
\end{figure*}

In Figure \ref{fig:fit_grid} (left), we show the $-{\rm ln}\mathcal{L}$-values, where dark colors indicate better fits to the control data for each velocity combination of the model streams in five of the ten different halos after the Jacobi radius cut (we skip every other halo in between for visualization purposes). 
We find that the Dw3 stream can not be reproduced in the lowest mass halo (Halo 1: M$_{200} = 0.94 \times 10^{12}$ M$_{\odot}$) since no  model stream covers the extent of the data after the Jacobi cut (see top row). Similarly, in Halo 2 (M$_{200} = 2.82 \times 10^{12}$ M$_{\odot}$), most of the stars are removed from the model streams after the Jacobi cut and only very poor fits to the data exist ($-{\rm ln}\mathcal{L} > 20$). Thus, the Dw3 stream can not be reproduced in the two lowest mass halos. 

For the rest of the halos shown in Figure \ref{fig:fit_grid} (left), the gray star marks the combination of $v_{\phi_1}, v_{\phi_2}$ that produced the best fit model stream in each halo.  We summarize the $-{\rm ln}\mathcal{L}$-values in each halo in Table \ref{tab:pot}.
The pockets of orbital solutions that produce good fits to the data in the different halos move to higher ($v_{\phi_1}, v_{\phi_2}$)-values for more massive halos, as expected from Figure \ref{fig:halomassdegeneracy}, since the progenitors must move faster to produce such a straight stream in a more massive potential. The exact combination of $v_{\phi_1}, v_{\phi_2}$ does not scale directly with halo mass (as in Figure \ref{fig:halomassdegeneracy}), since the radial velocity of Dw3, the Dw3 mass and the integration time are fixed in this experiment. Instead, the pockets of solutions shift to best reproduce the shape of the control data with a combination of $v_{\phi_1}, v_{\phi_2}$. 

In the second panel of Figure \ref{fig:fit_grid}, we show the model streams in $({\phi_1}, {\phi_2})$-space generated with the velocity combination that produced the best fit in the left panel. Since no model stream covered the extent of the data in Halo 1 after the Jacobi cut, we show an example of an orbit with ($v_{\phi_1}$,$v_{\phi_2}$) = (150,20) km s$^{-1}$.
The control data, $x_n$, are shown in black, while the simulated stream center points, $y_n$, are shown in gray.  Note that the gray center points for the simulated streams are difficult to see when they overlap well with the control data points in black.

In the third panel, we show the 3D physical separation between Dw3 and Cen A's center as a function of time. Note that the model stream for Halo 1 (top) is on first infall. The best fit stream in Halo 3 has only had one pericenter pass. In all other halos, Dw3 has completed two or more pericenter passes by the present day. 
In the right panels of Figure \ref{fig:fit_grid}, we show the best fit streams, but in ICRS coordinates projected onto the image data of Cen A and colored by the radial velocity in Cen A's rest frame (see color bar in Figure \ref{fig:fit_fixedhalo}). 

The lowest halo mass of Cen A in which we can fit the stream of Dw3 is Halo 3 (second row in Figure \ref{fig:fit_grid}) which has M$_{200}$ $ = 4.70\times 10^{12}$ M$_{\odot}$ (see also Table \ref{tab:pot}). However, the fit in this halo is poor ($-{\rm ln}\mathcal{L} = 7.83$) compared to the higher mass halos ($-{\rm ln}\mathcal{L} < 4$). 

To check if the poorer fit in Halo 3 is due to a resolution issue, we re-run a grid of orbits for Halo 3 stepping in increments of 2.5 km s$^{-1}$ in $v_{\phi_1}$ and 1 km s$^{-1}$ in $v_{\phi_2}$ instead. With the finer velocity resolution, we found a fit in Halo 3 that had $-{\rm ln}\mathcal{L} = 7.35$ where ($v_{\phi_1}, v_{\phi_2}$) = (345, 47) km s$^{-1}$. While this fit is worse than the fits for Halo 4 through 10 (see Table \ref{tab:pot}), we do not disregard this halo as a viable dark matter potential for Cen A.  

The lower limit of M$_{200} > 4.70\times 10^{12}$ M$_{\odot}$ (Halo 3) was constrained by only the one radial velocity measurement of Dw3 from \citet{dumont2022}. 
We notice, however, that the velocity gradient for a fixed distance from Dw3 also varies along the stream for the best fits (see colors in Figure \ref{fig:fit_grid} panels 2 and 4).

\begin{figure*}
\centerline{\includegraphics[width=\textwidth]{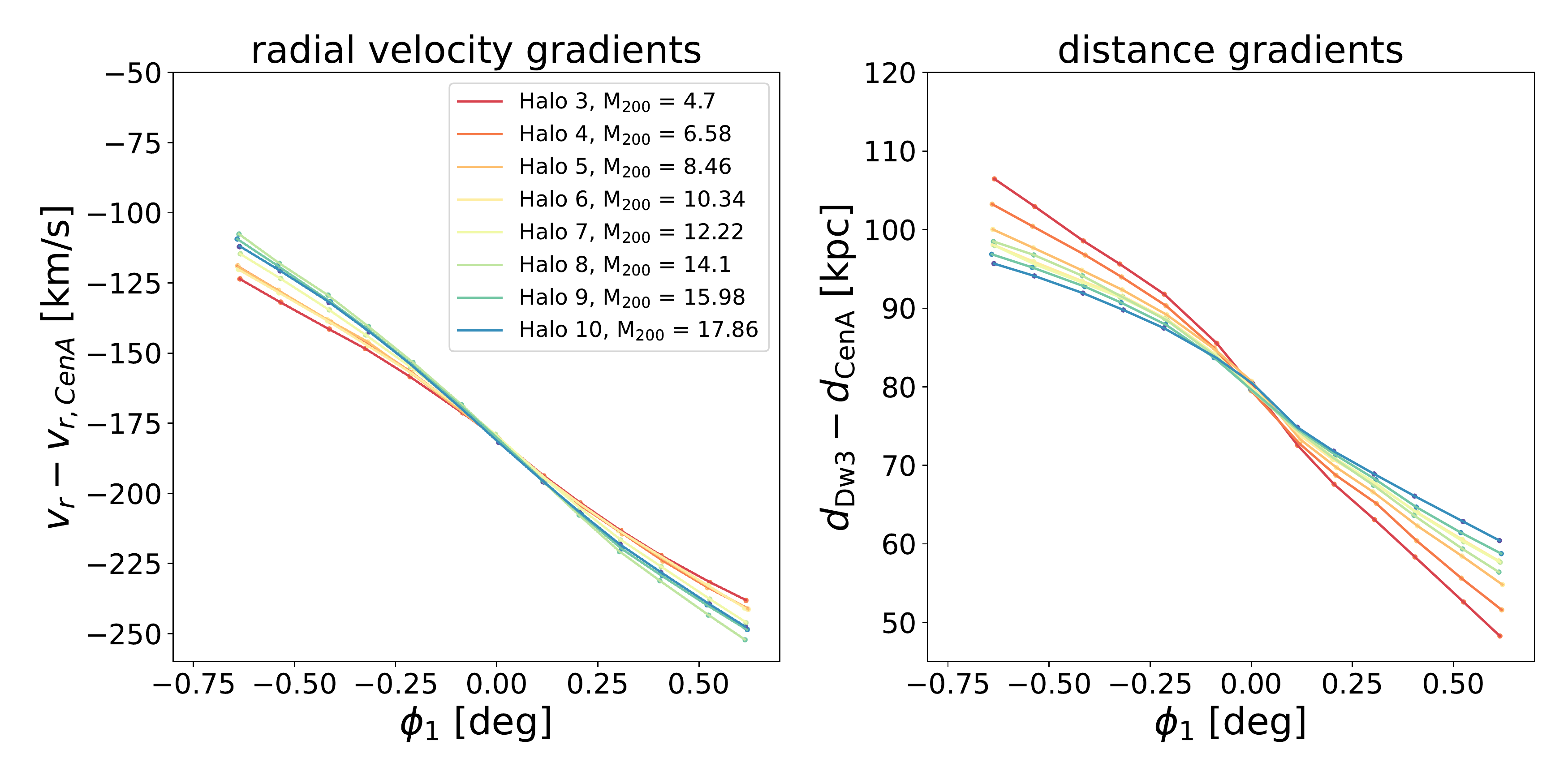}}
\caption{Left: Prediction of the technically observable radial velocity gradients for each best-fit streams in each of the 8 halos in which the stream could be well fit after applying the Jacobi criterion. All M$_{200}$ values are listed in units of $10^{12}$ M$_{\odot}$. The radial velocity is evaluated at the $\phi_1$ location of the 13 control data points, $x_n$ (see dots), for each best-fit model stream. The best fit model streams in the more massive halos have steeper velocity gradients.
Right: Same as the left panel but now for the line-of-sight distance gradient with respect to Cen A along each best fit stream in each of the 8 halos. The best fit model streams in the more massive halos have shallower distance gradients along the streams.}
\label{fig:vradgradient}
\end{figure*}

To explore this further, in Figure \ref{fig:vradgradient} we show the velocity gradients (left) and distance gradients (right) of each best fit simulated streams in the halos for which we could find a decent match to the control data (Halo 3 through Halo 10). In particular, we have evaluated the radial velocity and line-of-sight distance of the model stream simulations at the $\phi_1$ location of the 13 control points in the simulations and plotted a line through each point for each best fit model stream. 

The radial velocity gradients (left) all cross at $v_{rad} = -181.4$ km s$^{-1}$ since this velocity of Dw3 was fixed for all model stream simulations. Note how the radial velocity gradient along the stream gets steeper with higher halo masses (see blue line), as the stream needs to move faster to be on the same type of orbit. The radial velocity contrasts are larger at the edge of each stream, where the difference between Halo 3 (red; least massive) and Halo 10 (blue; most massive) is $\sim 10$ km s$^{-1}$. If we could measure radial velocities along the stream \citep[see e.g.,][]{toloba2016}, particularly at the end points of the streams, we could further constrain the dark matter halo mass.

The distance to Dw3 has been fixed to 3.88 Mpc and Cen A's distance has been fixed to 3.8 Mpc, thus all line-of-sight distances (Figure \ref{fig:vradgradient}, right) cross at a relative distance of 80 kpc. We find that the distance gradient is shallower in the highest mass halo (blue line), and that observations of distances at the end points of the streams could further constrain the dark matter halo mass. Both the radial velocity gradients and line-of-sight distance gradients are technically observable (see Section \ref{sec:distance} for a discussion of the feasibility of such measurements and their observational errors).

\section{Discussion}\label{sec:discussion}
We have shown that with just one radial velocity measurement of a stream progenitor combined with the sky track of an extragalactic stream, we can constrain properties of the dark matter halo of Cen A.  
In this Section, we discuss progenitor distance degeneracies (Section \ref{sec:distance}), as well as other degeneracies (Section \ref{sec:deg}), we compare our new Cen A halo constraints to other existing limits (Section \ref{sec:other}), we discuss the effect of including multiple Cen A streams in Section \ref{sec:multiple}, and we discuss the future of external galaxy stream science, and how to maximize the dark matter science output of these structures in Section \ref{sec:future}.

\subsection{Distance vs radial velocity degeneracy}\label{sec:distance}
If we did not know the distance to Dw3, we could produce a similar fit to the stream in the fiducial halo if we placed the dwarf in front of Cen A.
To demonstrate this, we first use the orbital solution of the best fit from Figure \ref{fig:fit_fixedhalo} in the fiducial halo (see Table \ref{tab:bestfit}),  
and create an array of distances to Dw3 from $-400$ to 400 kpc in increments of 15 kpc with respect to Cen A located at 3.8 Mpc.
We generate model streams where we place Dw3 at each of these distances and record the $-{\rm ln}\mathcal{L}$-value for each model stream compared to the data.

\begin{figure*}
\centerline{\includegraphics[width=0.8\textwidth]{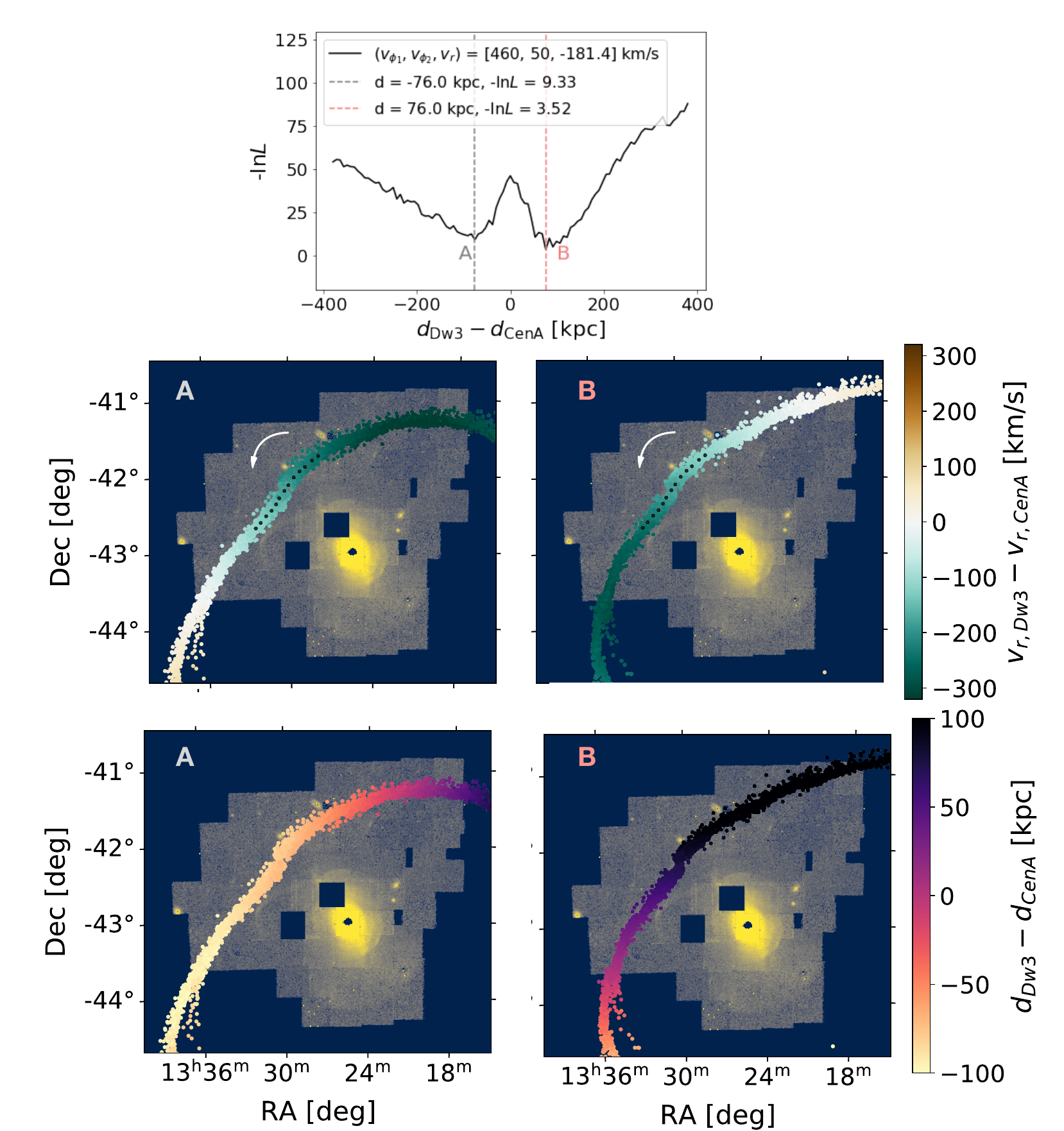}}
\caption{Top: $-{\rm ln}\mathcal{L}$ vs the progenitor's line-of-sight distance with respect to Cen A for model streams evolved for 6 Gyr in the fiducial potential with fixed radial velocity, progenitor mass and fixed ($v_{\phi_1},v_{\phi_2},v_r$)=($460,50,-181.4$) km s$^{-1}$ from the best fit model stream presented in Figure \ref{fig:fit_fixedhalo}. There are two clear minima (best fits), located in front of (A: $d = -76.0$ kpc) and behind (B: $d = 76.0$ kpc) Cen A (which is at 3.8 Mpc). 
Middle: Visualization of the model streams in ICRS coordinates from the best fit in front of (A: left) and behind (B: right) Cen A overplotted on the Cen A image data and colored by the radial velocity in Cen A's rest frame. Note how the radial velocities flip direction in the two cases. 
Bottom: same two model streams, but now colored by distance along the stream with respect to Cen A. The distance gradients in the southern part of the model streams are closer to us than the northern part in both cases. 
}
\label{fig:distance_degeneracy}
\end{figure*}

In the top panel of Figure \ref{fig:distance_degeneracy}, we plot the  $-{\rm ln}\mathcal{L}$-value as a function of Dw3 distance for each model stream. There are two clear minima (see labels A and B), which correspond to the two best fits to the control data points. The first minimum (A) is for Dw3 closer to us than CenA, at d = $-76$ kpc (see gray dashed line), and the second minimum (B) is for Dw3 farther from Cen A, at d = 76 kpc. In the two lower left panels, we visualize the best fit model stream in front of Cen A in ICRS coordinates overplotted on the image data of Cen A, where the stream is colored by the radial velocity along the stream in Cen A's rest frame (top) and colored by the distance gradient along the stream (bottom). In the lower right panels, we show the same plots, but for the Dw3 solution at a distance further away than Cen A. 

While the sign of the distance gradients are the same for the two fits (the southern parts of the stream are in the foreground of Dw3 in both cases), the sign of the radial velocity flips in the two cases. For case A, where Dw3 is closer to us than Cen A, the stream is moving towards us with respect to Cen A in the southern parts. For case B, where Dw3 is farther from us than Cen A, the stream is receding from us with respect to Cen A in the southern parts. Thus, if we can observe just the sign of the velocity gradient along the stream, this is a strong indicator of whether Dw3 resides behind vs in front of Cen A.

The model streams clearly predict that the northern part of the stream should be farther than Dw3 in both cases, and that the radial velocity gradient should be decreasing from north to south if Dw3 resides ``behind'' Cen A. 
Using the tip of the red-giant branch method, \citet{denja:2019} present distances along Dw3's stream from HST data (see their table 5). For three pointings, one placed on Dw3 and two placed in the northern and southern parts of the stream, respectively (see pointings in their Fig 13), the data from the ACS fields, show that the southern part is closer to us than Dw3, and that the progenitor distance and the northern stream distance are very similar (which resembles both of our cases in Figure \ref{fig:distance_degeneracy}). However, within the errors of their measurements, the three pointings are all consistent with being at the same distance. With larger field of views (e.g., with the Roman space telescope), however, a large fraction of the stream can be covered in one field, which might help determine continuous changes in distance along the stream and whether the northern part is in fact farther from us than Dw3. 

While it is encouraging that the best fit distance behind Cen A, d = 76 kpc, is close to the actual observed relative distance between Cen A and Dw3 of 80 kpc, recall that in the test in Figure \ref{fig:distance_degeneracy}, we have used the best fit orbit (see Table \ref{tab:bestfit}) for that specific d$_{\rm Dw3}$ = 3.88 Mpc distance. Thus, the fact that we obtain a similar distance is not surprising. Similarly, it is not surprising that Case B produces a better fit ($-{\rm ln}\mathcal{L}_B =  3.52$) than Case A ($-{\rm ln}\mathcal{L}_A =  9.33$) for that same best fit orbit. When we run a grid of orbits for model streams where Dw3 is placed in front of Cen A (at d = $-76$ kpc with respect to Cen A), we find just as good fits to the data as in Case B. 
The radial velocity gradients in the fits where Dw3 is behind vs in front of Cen A have the opposite sign, while the distance gradients have the same sign (as expected from Figure \ref{fig:distance_degeneracy}). We emphasize again how an observation of just the sign of the velocity gradient is very informative.

\subsection{Other degeneracies and limitations of our method}\label{sec:deg}
Throughout this work, we have fixed certain parameters. Here, we discuss how these decisions affect our results. 

\begin{itemize}
\item[]{\bf Integration time}: When we evaluate the fits to the data, we have discarded model streams that do not cover the extent of the 13 control points after 6 Gyr of evolution. However, if we integrate some of those disregarded streams for a longer period, it is possible that they could cover the extent of the data\footnote{Note that some model streams point in the radial direction and that a longer integration time would not help in these cases.}. To check how different integration times affect our results, we re-run the grid of ($v_{\phi_1}$,$v_{\phi_2}$)-velocities in Figure \ref{fig:orbits}, but integrate the model stream orbits and release stream stars for 4 Gyr and 8 Gyr instead of 6 Gyr.
We find that some of the regions with high $-{\rm ln}\mathcal{L}$-values (see color bar in Figure \ref{fig:orbits}) move towards lower values (better fits) for the 8 Gyr integration time, and some of the regions with low $-{\rm ln}\mathcal{L}$-values (see color bar in Figure \ref{fig:orbits}) move towards higher values (worse fits) for the 4 Gyr integration time. 
However, this does not change our main best-fit orbital solution. A more important factor than the integration time is the fraction of the orbital time during which the progenitor can actually tidally strip stars (see Eq. \ref{eq:jacobi}). 

\item[]{\bf Progenitor mass:} In Figure \ref{fig:halomassdegeneracy}, we showed that if we simulated Dw3 in a lower mass halo, but integrated its orbit for longer with a lower progenitor mass and orbital velocity, we could reproduce the same morphology of the stream as in a more massive halo. Thus, if we lower the mass of the progenitor, lower the orbital velocity but prolong the integration time, we could potentially produce similar solutions to our best fit. However, since we have a radial velocity measurement of Dw3, we cannot arbitrarily scale down the orbital velocity. 
A good constraint on the progenitor mass does help break degeneracies on integration time, and also affects the Jacobi radius in a specific halo (see Eq. \ref{eq:jacobi}). If the progenitor had a higher initial mass, some of which can be in the form of dark matter (see Section \ref{sec:nbody}), it would be harder to strip stars from the dwarf. A more massive Dw3 would need to be evolved in an even higher Cen A halo mass to have its stars stripped. Thus the lower limit presented in this work would only become even more constraining in that case. A higher progenitor mass would also affect the width of the stream, which has been left as a free parameter in this work. 

\item[]{\bf Cen A distance:} 
Throughout this work, we have fixed the distance to Cen A to 3.8 Mpc, which sets the physical scale of our coordinate system. If Cen A instead was closer to us, this would lead to a smaller physical projected separation between Dw3 and Cen A (instead of 79 kpc). If we use the same best-fit velocity as in Figure \ref{fig:fit_fixedhalo}, Dw3 would complete more orbits, reach a smaller pericenter, and the tidal radius would therefore also be smaller. This would, in turn, change our limits on the minimum halo mass. However, the error on the distance estimate for Cen A for \citet{Harris2010} is $d = 3.8~\pm~0.1$ Mpc, so this is a minimal effect, although we might have larger distance uncertainties for smaller dwarf galaxies or more distant, less studied galaxies in future analyses. 

\item[]{\bf NFW potentials:}
We primarily use the fiducial potential (see Section \ref{sec:potential}) motivated by enclosed mass estimates from GCs, and also explore a range of 10 potentials with $[0.1 - 1.9] \times$ M$_{\rm 200,fiducial}$. In this work,  the lower limit on the dark matter halo mass for Cen A is determined from the minimum halo mass of those ten potentials in which the straight Dw3 stream's control data could be reproduced. However, an exploration of the lower limit through a finer range of NFW potentials could yield a better lower limit. Given the limitations listed above, we do not explore more potentials in this work. 

\item[]{\bf Time-dependence}: We have assumed that the potential of Cen A is static. From recent work on stellar streams in the Milky Way \citep[e.g.,][]{shipp2021,Lilleengen2022,dillamore2022}, we know that both thin GC streams and dwarf galaxy streams are sensitive to perturbations from accreted dwarf galaxies \citep[see also the LMC's affect on the Milky Way's dark matter halo in][]{Garavito2019}. \citet{nibauer2022} argue that not all streams are sensitive to such perturbations (depending on the streams' orbits with respect to the infalling satellite's orbit), and that some streams can still provide powerful constraints on dark matter halo parameters despite time-dependence. It is unclear how much the assumption that Dw3 evolves in a static potential for 6 Gyr used in this work affects our results. 
 
\end{itemize}

\subsection{Comparison to other dark matter halo mass limits}\label{sec:other}
After exploring a range of NFW potentials to represent Cen A's dark matter halo, we found that the straight stream from Dw3 can be reproduced in a potential with an NFW profile with M$_{200} > 4.70 \times 10^{12}$ M$_{\odot}$. 
Using 3D distances, line-of-sight velocities, and a Bayesian model to study the kinematics of the
satellite systems of Cen A, \citet{mueller2022} derive
M$_{200} = 5.3 \pm 3.5 \times 10^{12}$ M$_{\odot}$, similar to our limit within the errors. 

\citet{peng2004} reported Cen A's enclosed mass at 40 kpc ($\sim 36\arcmin$) using 215 GCs to be $M(R<40~ {\rm kpc}) = 7.5 \times 10^{11}$ M$_{\odot}$ \citep[see ][for this corrected mass]{woodley2007}. For comparison, \citet{Woodley2010} set a limit on the enclosed mass of $M(R<40~ {\rm kpc}) = 9.7~\pm~3.3~\times~ 10^{11}$ M$_{\odot}$ using 429 GCs in Cen A. In this work the potential with the minimal halo mass (Halo 3) that could reproduce the straight stream from Dw3 has $M(R<40~ {\rm kpc}) = 7.21 \times 10^{11}$ M$_{\odot}$ (see Table \ref{tab:pot}). Overall, the lower limit on the dark matter halo mass provided in this paper is in agreement with previous studies.

\subsection{What about Cen A's other substructure?}\label{sec:multiple}
In addition to the stream emerging from Cen A, there are several shell features and a prominent ``second stream'' (see Figure \ref{fig:cena_data}) present in the data. \citet{Wang2020} showed that a single major merger event 2 Gyr ago with a mass ratio of up to 1.5 can account for much of the shell structure in the halo, and in some cases produce features mimicking the Dw3 stream and the ``second stream''. In this section, we investigate the ``second stream'' as an individual stream originating from a dwarf galaxy progenitor.
\citet{denja:2016} find that the second stream is at the same distance as Cen A. For the remainder of this Section, we place both Cen A and the second stream progenitor at 3.8 Mpc.
There is no public radial velocity estimate, but  \citet{Hughes2022} finds a GC at the center of the stream with $v_{\rm rad} = 651 \pm 13$ km s$^{-1}$, i.e., redshifted by 110 km s$^{-1}$ with respect to Cen A.

We can easily, and not surprisingly, find an orbit that reproduces the morphology of the second stream in the Cen A fiducial halo (black line in Figure \ref{fig:potential}). To search for the best fit, we first place three control data points along the stream. We assume that the progenitor is at the center of the stream, and place two other control points in the leading and trailing arm to evaluate how well the model streams fit the data (see white points in Figure \ref{fig:cena_data}). Note that it is unclear which arm is trailing and leading. We fix the radial velocity to $v_{\rm rad}$ = 110 km s$^{-1}$ in Cen A's rest frame. 
We again define a new coordinate system (here $\psi_1,\psi_2$ in degrees). We first shift the coordinate system such that Cen A is at $(\psi_1,\psi_2) = (0,0)^\circ$, and then rotate the coordinate system by $82^\circ$, such that the progenitor is located at $\psi_1 = 0$, directly above Cen A. We again use the \texttt{Astropy} coordinate transformation machinery \citep{astropy13,astropy18} to transform the ICRS coordinates (see Table \ref{tab:cena}) of the second stream to kpc for stream modeling. 

We fix the distance to the progenitor (see Table \ref{tab:cena}), and explore a grid of velocities in the $v_{\psi_1}$ and $v_{\psi_2}$ direction of the second stream. We first integrate the progenitor orbit for 6 Gyr for each orbit combination on a velocity grid ranging from $-300$ to 300 km s$^{-1}$ in steps of 25 km s$^{-1}$ in both $v_{\psi_1}$ and  $v_{\psi_2}$. There is a pocket of solutions which has the stream moving with either both positive or both negative $v_{\psi_1}$ and $v_{\psi_2}$ combinations. To resolve the best fit, we run a zoomed version of the peak near $v_{\psi_1}$= 150 -- 250 km s$^{-1}$ in steps of 10 km s$^{-1}$ and $v_{\psi_2}$ = 40 -- 100 km s$^{-1}$ in steps of 2 km s$^{-1}$. Note that no stars are removed with the Jacobi cut for these orbits in this fiducial potential.

\begin{figure*}
\centerline{\includegraphics[width=\textwidth]{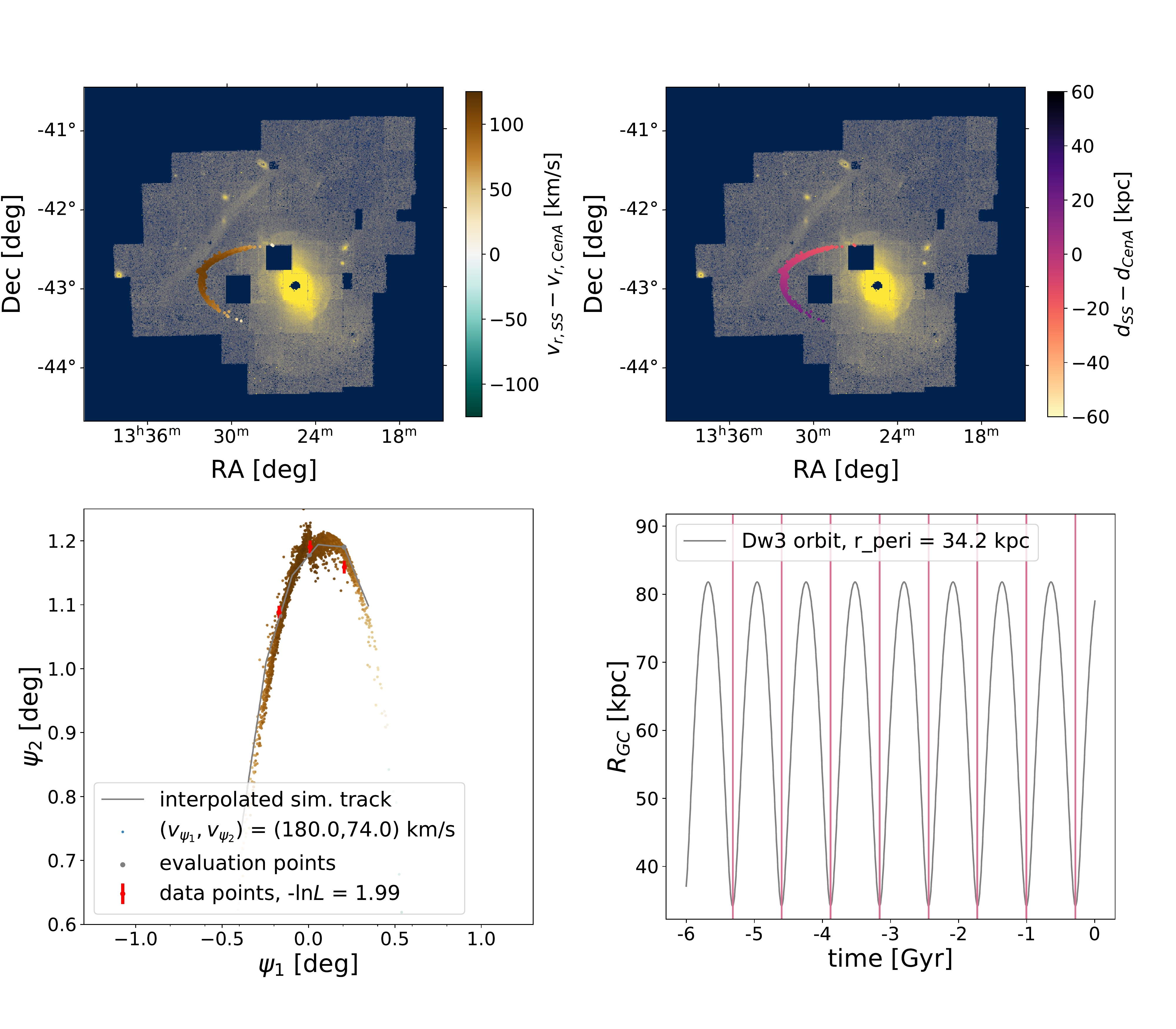}}
\caption{The best fit to the three control data points of the ``second stream'' (SS) in Cen A's halo evolved in the fiducial potential integrated for 6 Gyr. 
We fix the radial velocity to 110 km s$^{-1}$ with respect to Cen A's rest frame and find that ($v_{\psi_1},v_{\psi_2}$) = (180,74) km s$^{-1}$ minimizes $-{\rm ln}\mathcal{L}$ to be 1.99.
Upper panels: ICRS projection of the best fit simulated stream overplotted on the Cen A image data colored by the radial velocity in Cen A's rest frame (left) and the distance along the stream with respect to Cen A's distance (right). 
Bottom left: the best fit stream visualized in ($\psi_{1},\psi_{2}$)-space. The red points are the three control data points. The grey points are the center points of the simulated stream evaluated at the $\psi_{1}$-position of the control data points. 
Bottom right: galactocentric 3D position of the simulated progenitor as a function of time. 
}
\label{fig:stream2}
\end{figure*}

The best fit is for ($v_{\psi_1}, v_{\psi_2}$) = (180, 74) km s$^{-1}$, where $-{\rm ln}\mathcal{L}$ = 1.99, which we present in Figure \ref{fig:stream2}. Note that there was a very similar fit where the stream moved in the negative ($v_{\psi_1},v_{\psi_2}$)-direction.
In the top panels we show the stream projected in ICRS coordinates overplotted on the Cen A image data colored by the second stream's simulated radial velocity gradient in Cen A's rest frame (left) and distance gradient with respect to Cen A (right). Our best fit model stream shows a very flat velocity gradient, where the ends of the stream are slightly less redshifted with respect to Cen A. 
We plot the model stream in the $\psi_1, \psi_2$-space (lower left), as well as the evolution of the 3D position of the progenitor with respect to the center of Cen A (lower right). The progenitor is close to apocenter today. We find a very similar best fit in the minimum halo mass that could reproduce the Dw3 stream (Halo 3: M$_{200}$ = $4.7\times10^{12}$ M$_{\odot})$: ($v_{\psi_1}, v_{\psi_2}$) = (150, 64) km s$^{-1}$, with $-{\rm ln}\mathcal{L}$ = 2, as well as in the lowest mass halo (Halo 1: M$_{200}$ = $9.4\times10^{11}$ M$_{\odot})$: ($v_{\psi_1}, v_{\psi_2}$) = (90, 44) km s$^{-1}$, with $-{\rm ln}\mathcal{L}$ = 2.32.

\citet{Bonaca2018} found that multiple streams in one halo have greater constraining power on dark matter halos. We plan to extend the method of \citet{Bonaca2018} to external galaxies, and explore the information content of multiple extragalactic streams observed in projection. 
We leave any further exploration, particularly on the second stream and the Dw3 stream's combined constraining power on the Cen A halo concentration and shape, for future work.

\subsection{The future of extragalactic streams for dark matter science}\label{sec:future}
Extragalactic stellar streams offer one of the few methods for measuring the dark matter halos in individual galaxies. Observing radial velocities of low surface brightness streams is tricky, but 
crucial, if we want to use the many future stream detections (e.g., from Roman, Rubin, and Euclid) 
for dark matter science. In this work, we have used the radial velocity estimate from the nuclear star cluster in Dw3 \citep{dumont2022}. But often we do not detect the progenitors of extragalactic streams \citep[see e.g.,][]{delgado2021} and can only observe the diffuse light in the actual streams. 

Even without a known progenitor, like the Dw3 nuclear star cluster used in this work, GCs provide bright signposts that can be used to determine stream velocities. Significant GC populations are found associated with MW streams \citep[e.g.,][]{malhan2022} as well as those in M31 \citep{Veljanoski2014}. 
With the caveat of not always knowing the line-of-sight distance, and thereby the true association of the GCs with the streams, velocities of GCs can be measured with current instrumentation out to distances of $\sim$30 Mpc \citep[][]{brodie2014}. Future extremely large telescopes will extend this distance limit by about a factor of 3, potentially enabling halo mass determinations in a wide range of halo masses and environments with the method presented in this paper.

If the streams do not host GCs, the radial velocities along the streams can also be estimated from the technique presented in  \citet{toloba2016}. In particular,  they study the internal stellar kinematics of low surface brightness streams beyond the Local Group. Their technique relies on  multi-object spectroscopy of individual stars using the DEIMOS spectrograph \citep{Faber03} on Keck \citep[see also][for DEIMOS spectra of individual stars in the Giant Southern Stream of M31]{raja2006}, although similar science can be done with any large ground-based telescope with wide-field multi-object spectroscopy capabilities. They place slitlets on the brightest available stars in the stream, including tip of the red giant branch and asymptotic giant branch stars, along with stellar blends.  
By using traditional spectral stacking techniques \citep[see][Section 2 for a detailed description]{toloba2016}, sufficient signal to noise is built up to measure bulk kinematic properties.
\citet{toloba2016} apply their method to the stream around the NGC 4449 dwarf galaxy, which is at a distance of 3.82 Mpc \citep{delgado2012}, and find a flat velocity gradient along the NGC 4449 stream with a typical error of 10--20 km s$^{-1}$. 
See also \citet{toloba16b} which used this method to determine the velocity and metallicity of an M81 satellite dwarf galaxy.
Cen A is at a similar distance as NGC 4449, and we expect similar errors for any future radial velocity measurements along the Dw3 stream with this method. Based on Figure \ref{fig:vradgradient}, we note that while this will only allow us to, at best, distinguish between the most extreme cases of viable dark matter halos, even just the sign of the radial velocity gradient along the stream will be informative.  

While DEIMOS can only be used out to $\sim$5 Mpc in the northern sky \citep{toloba16b}, it is possible that the Very Large Telescope's (VLT's) IFU can be used for similar purposes. 
ESO's Extremely Large Telescope (ELT, planned to receive first light in 2027) will be suitable for the \citet{toloba2016} method, and could potentially target faint surface brightness features, such as streams, out to distances 2--3 times farther than the VLT (e.g., the Virgo Cluster at $d\sim 16.5$ Mpc). 

While we might not be able to obtain radial velocities along all future observations of extragalactic streams, even with the ELT or GCs, these streams can still inform accretion histories and their number counts and surface brightness will provide crucial comparisons to $\Lambda$CDM (e.g., Starkenburg et al., in prep.). 

\section{Conclusion}\label{sec:conclusion}
In this paper, we have explored whether the straight stream emerging from Dw3 could be reproduced in the halo of the elliptical galaxy Cen A, and what constraining power the stream holds on Cen A's dark matter halo mass. We have used simple stream models and a new external galaxy stream-fitting technique to explore large grids of orbital parameters. We combined our method with a Jacobi radius cut to ensure realistic tidal stripping. We conclude the following:

\begin{itemize}
    \item To reproduce the straight nature of the Dw3 stream at a projected distance of 79 kpc in Cen A's halo, we find that only a small pocket of orbital solutions for the Dw3 progenitor yield good fits to the data. These solutions all require that Dw3 moves very rapidly in the direction parallel to the stream and slowly in the direction perpendicular to the stream. 
    
    \item If we only have observations of the morphology and track of a stream, it is not possible to constrain the dark matter halo, since the orbital velocity of the stream, as well as the progenitor mass and integration time can be scaled to produce the exact same morphology in an arbitrary halo mass. 
    
    \item If we include observations of just one radial velocity, we can constrain the halo mass of Cen A. When we fix the observed distances to Cen A and Dw3, the Dw3 stream can only reproduce the control data for a minimum Cen A halo mass of M$_{200} > 4.70 \times 10^{12}$ M$_{\odot}$. This is consistent with studies of Cen A's GC and dwarf galaxy kinematics. 
    
    \item We test the best fit model stream's morphology with an N-body simulation including a Dw3 dark matter component, and confirm that we obtain the same stream morphology with more realistic tidal stripping conditions. 
    
    \item The radial velocity and distance gradients along the Dw3 stream can help constrain the halo mass of Cen A further. In particular, the end points of the stream are most informative, since they display the largest difference for different halo masses.
    
    \item If the line-of-sight distance to Dw3 is greater than the line-of-sight distance to Cen A, our models predict that the southern parts of the Dw3 stream are more blueshifted than the northern parts. The sign of the radial velocity gradient along the stream can help distinguish between whether Dw3 is located in front of or behind Cen A. 
    
    \item The second stream in Cen A's halo can easily be reproduced in the fiducial halo of Cen A as well as in the lowest mass halo included in this work.
\end{itemize}

We stress that extragalactic stellar streams provide a fundamental way of mapping the dark matter in our Universe, which is otherwise only possible through lensing studies, GC kinematics, integrated light velocity dispersions, and rotation curves of galaxies. Given the streams' potential to make discoveries about dark matter, we urge the extragalactic stream community to focus their efforts on radial velocity follow-up plans.\\

\acknowledgements
We thank the CCA Dynamics Group, E. Tollerud, A. Bonaca, A. Dumont, J. Strader, and T. Starkenburg for insightful discussions. We also thank J.~Strader and A.~K. Hughes for providing Centaurus A globular cluster data and analysis.
Support for this work was provided by NASA through the NASA Hubble Fellowship grant \#HST-HF2-51466.001-A awarded by the Space Telescope Science Institute, which is operated by the Association of Universities for Research in Astronomy, Incorporated, under NASA contract NAS5-26555. 
The Flatiron Institute is supported by the Simons Foundation. DJS acknowledges support from NSF grants AST-1821967 and 1813708. Research by DC is supported by NSF grant AST-1814208.

\software{
   \package{Astropy} ~\citep{astropy13,astropy18}, 
    ~\package{Matplotlib} ~\citep{Hunter:2007}, 
    ~\package{Gala} ~\citep{gala2017,gala2020}, 
    ~\package{Numpy} ~\citep{walt2011}, 
    ~\package{Scipy} ~\citep{scipy}. 
}


\end{document}